# P vs NP Problem in Portfolio Optimization: Integrating the Markowitz–CAPM Framework with Cardinality Constraints and Black–Scholes Derivative Pricing

**Davit Gondauri,** ORCID: https://orcid.org/0000-0002-9611-3688
Professor, Doctor of Business Administration, Business & Technology University, Georgia
**Corresponding author:** Davit Gondauri, Dgondauri@gmail.com
**Type of manuscript:** research paper

Abstract

This study develops an integrated economic–computational framework for portfolio construction that makes the P versus NP divide operational within a financially auditable Markowitz–CAPM setting. Starting from the convex mean–variance program, we impose a hard cardinality constraint $|supp(w)| \leq K$, which couples discrete subset selection with continuous weight optimization and yields a mixed-integer quadratic program (MIQP). To ensure empirical transparency, the asset universe is built from approximately n≈94 U.S. industry portfolios from Aswath Damodaran, using levered betas and annualized equity volatilities to calibrate expected returns via $\mu\_i = R\_f + \beta\_i \cdot ERP$ and to construct a fully reconstructible covariance matrix through a single-index (market-model) structure. Because the resulting search space grows combinatorially (≈C(n,K)), the paper treats scalable optimization as an approximation problem and evaluates practical solution schemes—greedy screening, Monte Carlo sampling over K-subsets, and genetic algorithms—under a replication-oriented protocol with random-seed logging, distributional performance reporting (median/IQR/quantiles), convergence/effort curves, and runtime profiling. Dependence diagnostics (median pairwise correlation, tail correlation shares, eigenvalue concentration, PSD checks) are reported as first-class outputs to connect algorithmic behavior to the geometry implied by Σ. To support financial realism beyond linear assets, the framework is extended to a derivative-augmented universe by embedding a Black–Scholes European call as an additional instrument, mapped into CAPM-consistent moments via delta-based linearization and validated by moneyness–maturity robustness and a bump test (delta vs. repricing). Results demonstrate that (i) the hard cardinality rule materially reshapes attainable efficient frontiers and induces discontinuities relative to the unconstrained benchmark; (ii) heuristic performance must be assessed by stability and compute-cost trade-offs rather than single best-run outcomes; and (iii) under a single-index covariance, strong common-factor dependence limits diversification, explaining why high-β industries and clustered sectors may co-appear in K-sparse solutions. Overall, the paper provides a journal-grade, reproducible template for studying NP-hard portfolio selection with transparent inputs and extensible derivative overlays.

## 1. Introduction

Modern portfolio construction sits at an uncomfortable but intellectually productive intersection of economics, mathematics, and computation. At the economic layer, the Markowitz mean–variance paradigm and CAPM-style expected return calibration offer a disciplined way to map trade-offs between risk and reward. At the computational layer, real-world mandates—limited position counts, budget and concentration limits, and derivative overlays—convert an otherwise convex optimization into a discrete–continuous problem whose search space grows combinatorially. This paper uses that tension as its organizing principle: it treats cardinality-constrained portfolio selection as a concrete instantiation of the P vs NP divide, and then demonstrates how a journal-grade empirical workflow can be built around it—transparent inputs, auditable covariance construction, reproducible stochastic search, and diagnostics that connect algorithmic behavior to the underlying dependence structure.

The core difficulty is structural. In the classical unconstrained Markowitz program, feasible sets are convex and the efficient frontier can be computed via polynomial-time convex optimization. The moment a hard cardinality rule is imposed—|supp(w)| ≤ K with binary selection variables—portfolio choice becomes a mixed-integer quadratic program (MIQP): a coupling of discrete subset selection and continuous weight optimization. This coupling aligns portfolio selection with the broader family of sparse quadratic programs and subset-selection problems that are widely recognized as computationally intractable in the worst case. In complexity terms, candidate solutions can be verified efficiently, but globally optimal support discovery may require exploring up to C(n,K) subsets. As a result, tractable practice relies on approximation—heuristics and metaheuristics—whose credibility must be earned through reproducibility, convergence profiling, and stability reporting rather than through optimality guarantees.

A second motivation is financial realism. Institutional portfolios rarely consist of "infinitesimal" allocations to hundreds of names; they are implemented as sparse books with monitoring, liquidity, and governance constraints. Moreover, modern portfolios often include nonlinear instruments. Options, in particular, are not merely side products: they are explicit mechanisms for convex payoffs, tail shaping, and risk transfer. Yet the presence of derivatives raises a methodological question: how can one embed option exposures into a mean–variance scaffold without collapsing auditability? The present study answers with a replicable, delta-based embedding of a European call priced by Black–Scholes, mapped into CAPM-consistent moments and integrated into the same covariance construction used for the base universe. The point is not to claim a complete derivatives engine; it is to show a reproducible bridge from a nonlinear payoff to an augmented mean–variance instance that can be analyzed under the same cardinality logic.

### 1.1 Research question and positioning

The research question is deliberately operational: given an empirically grounded universe of industry portfolios, how does the computational hardness induced by a hard cardinality constraint shape (i) attainable efficient frontiers, (ii) algorithm choice and convergence behavior, and (iii) the interpretability of the resulting portfolios—including when the universe is augmented with a derivative instrument? By coupling a standard financial scaffold (Markowitz–CAPM) with explicit complexity framing, the paper provides an integrated narrative that is often treated in fragments: portfolio theory on one side, and algorithm engineering on the other.

### 1.2 Empirical design and data provenance

To ensure that results are interpretable and reproducible, the empirical universe is built from Aswath Damodaran's industry portfolios (approximately n≈94), using levered betas and annualized equity volatilities as the primary inputs for expected return calibration and covariance construction. Expected returns are generated by a CAPM mapping $\mu\_i = R\_f + \beta\_i \cdot ERP$, while the covariance matrix $\Sigma$ is constructed via a market-model (single-index) structure that is consistent with the beta inputs and matched to each industry's total variance. This design choice is intentional: it makes the full $\Sigma$ and $\rho$ matrices reconstructible from an auditable input table even when raw return series are not carried in the manuscript. At the same time, it imposes a strong common-factor geometry on dependence; therefore the paper treats correlation diagnostics (median pairwise correlation, tail shares, eigenvalue concentration) as first-class outputs rather than as afterthoughts.

**1.3 Contributions**

The paper makes five contributions that, taken together, aim to meet reviewer expectations for both theoretical anchoring and empirical transparency:

- Complexity anchoring: the cardinality-constrained mean–variance program is formalized with an explicit decision version and positioned within NP-complete/NP-hard taxonomy, clarifying why exact global optimization is not expected to scale in n and K.
- Replicable empirical backbone: all portfolio metrics ($\mu\_p$, $\sigma\_p$, Sharpe) are computed from a documented CAPM calibration and a fully specified $\Sigma$ construction, enabling independent reconstruction and audit.
- Heuristic evidence beyond "best run": stochastic methods are reported with distributional statistics (median, IQR, quantiles) across multiple random seeds, alongside convergence/effort curves and runtime profiling.
- Dependence diagnostics for interpretability: correlation and covariance summaries and heatmaps are treated as explanatory objects that connect algorithm outputs (selected subsets) to the geometry implied by the covariance model.
- Derivative integration as a reproducible overlay: a Black–Scholes European call worked example is embedded into the same Markowitz–CAPM instance via delta-based linearization, with moneyness–maturity robustness and a bump test to quantify approximation error.

**1.4 Organization of the paper**

The remainder of the manuscript proceeds as follows. Section 3 formulates the Markowitz–CAPM + cardinality MIQP, documents the Damodaran-based calibration, and specifies the algorithms and reproducibility protocol. Section 4 reports empirical frontiers, portfolio compositions, distributional Sharpe diagnostics, dependence summaries, and computational profiling; it then extends the analysis to derivative-augmented cases and approximation checks. Throughout, every table and figure is paired with a source note and parameter disclosure so that computational claims can be verified independently.

## 2. Literature Review

**2.1 Foundations of portfolio theory: mean–variance efficiency**

Modern quantitative portfolio selection begins with the mean–variance paradigm, in which the investor chooses portfolio weights to trade expected return against variance as a proxy for risk. The seminal contribution formalizes diversification, the efficient frontier, and tangency-portfolio logic under quadratic risk and linear return aggregation (Markowitz, 1952, 1959).

Later syntheses clarify that mean–variance efficiency is exact under elliptically distributed returns (or quadratic utility) and becomes an approximation otherwise; nevertheless, it remains the canonical benchmark for constrained allocation studies because the objective and constraints are explicit, auditable, and computationally tractable in its unconstrained convex form (Bodie et al., 2021; Elton et al., 2014; Boyd & Vandenberghe, 2004).

The unconstrained Markowitz program is solvable via convex quadratic programming, with a continuous feasible set that yields a smooth frontier parameterized by a risk-aversion scalar or by a target-return grid. Convexity is central: it is what makes efficient-frontier computation polynomial-time in the classical setting (Boyd & Vandenberghe, 2004; Nesterov & Nemirovskii, 1994).

## 2.2 Expected return calibration: CAPM and modern asset-pricing perspectives

To operationalize mean–variance optimization, expected returns must be specified. The capital asset pricing model (CAPM) provides a parsimonious mapping from systematic risk exposure to expected excess return, enabling transparent calibration of $\mu$ through $\beta$-loadings and an equity risk premium (Lintner, 1965; Mossin, 1966; Sharpe, 1964).

Empirical critiques and extensions highlight that single-factor CAPM is often incomplete in explaining the cross-section; multi-factor models (notably size and value) and broader asset-pricing frameworks provide alternative risk-premium decompositions and may materially change implied $\mu$ vectors (Cochrane, 2005; Fama & French, 1993, 2004).

In this document's design, CAPM is used as an auditable, reconstructable prior for $\mu$, not as a claim of realized performance. This aligns with applied optimization practice: use a disciplined parametric return model to isolate computational and constraint-induced geometry, then subject conclusions to sensitivity tests on premium and rate inputs (DeMiguel et al., 2009; Fama & French, 2004; Elton et al., 2014).

## 2.3 Risk modeling and dependence structure: covariance estimation and factor models

Covariance estimation is the second pillar of mean–variance analysis. Classical implementations rely on sample covariances, but finite-sample error can destabilize optimized weights; this motivates shrinkage estimators, Bayesian adjustments, and factor-structure approaches that trade bias for variance reduction (Jorion, 1986; Ledoit & Wolf, 2004).

The single-index (market) model provides an analytically convenient factorization, decomposing each asset return into a market-driven component and an idiosyncratic residual. This yields a covariance structure with rank-one commonality plus diagonal idiosyncratic terms—attractive when full return histories are unavailable but $\beta$ and $\sigma$ inputs are (Sharpe, 1963; Elton et al., 2014).

While factor models enhance interpretability, they can compress the effective diversification space when the common factor dominates. Dynamic correlation models and multi-factor risk models provide richer dependence representations and are frequently used as robustness alternatives when single-index structure may over-impose comovement (Campbell et al., 1997; Engle, 2002).

## 2.4 Real-world constraints and the emergence of mixed-integer portfolio optimization

Institutional portfolios face constraints absent from the idealized Markowitz program: limited position counts (cardinality), buy-in thresholds, minimum/maximum holdings, sector caps, turnover bounds, and discrete-lot trading constraints. These restrictions are empirically motivated—monitoring capacity, liquidity, transaction costs, governance—but they fundamentally alter the optimization landscape (Jobst et al., 2001; Mansini & Speranza, 1999; Schaerf, 2002).

A hard cardinality constraint, $|\text{supp}(w)| \leq K$, introduces binary selection variables coupled to continuous weights, producing a mixed-integer quadratic program (MIQP). This discrete–continuous coupling breaks convexity and makes the efficient frontier discontinuous under many practical restrictions. Consequently, the literature develops exact methods (branch-and-bound/cut) and heuristic approximations (Bertsimas & Shioda, 2009; Chang et al., 2000; Gao & Li, 2013).

## 2.5 Computational complexity: P vs NP framing, NP-hardness, and sparse quadratic programs

The P vs NP framework distinguishes problems whose candidate solutions can be verified efficiently from problems whose optimal solutions are not known to be computable in polynomial time. Complexity theory formalizes NP-completeness and NP-hardness and provides reduction techniques used to position applied optimization tasks (Arora & Barak, 2009; Cook, 1971; Garey & Johnson, 1979; Karp, 1972; Papadimitriou, 1994).

Cardinality-constrained mean–variance optimization is closely related to sparse quadratic programming and subset-selection problems: one must discover a support set of size K that jointly optimizes a quadratic form, which in the worst case requires combinatorial exploration. In the broader optimization and statistics literatures, best-subset selection is NP-hard, and mixed-integer optimization has become a primary tool for obtaining exact or certifiably near-exact solutions in moderate-scale instances (Bertsimas et al., 2016; Del Pia et al., 2020; Natarajan, 1995; Tibshirani, 1996).

For portfolio optimization, discrete asset-choice constraints therefore shift the problem class away from convex QP into NP-hard MIQP/QMIP territory. Large-n empirical work typically pairs this complexity framing with approximation schemes and transparent convergence evidence rather than claiming global optimality at scale (Bertsimas & Shioda, 2009; Chang et al., 2000; Jobst et al., 2001).

**2.6 Exact algorithms for cardinality-constrained portfolio selection**

Exact approaches treat cardinality-constrained mean–variance selection as MIQP/QMIP and leverage branch-and-bound, cutting planes, and Lagrangian relaxations. Early branch-and-cut studies demonstrated the feasibility of solving meaningful instances by exploiting problem structure, while later contributions improved formulations, bounds, and decomposition to expand solvable sizes (Bienstock, 1996; Cesarone et al., 2013; Gao & Li, 2013; Shaw et al., 2008).

Exact methods remain essential as benchmarking instruments: reduced-universe instances provide ground-truth optima against which heuristic gaps can be quantified. This practice aligns with general integer-optimization methodology, where exact solvers deliver certificates and support controlled algorithmic experiments (Bertsimas & Tsitsiklis, 1997; Nemhauser & Wolsey, 1988; Bertsimas & Shioda, 2009).

**2.7 Heuristics and metaheuristics: genetic algorithms, simulated annealing, tabu search, and Monte Carlo sampling**

Given combinatorial growth in C(n,K), heuristics and metaheuristics are widely used. Canonical portfolio studies compare genetic algorithms (GA), tabu search, and simulated annealing (SA) for approximating cardinality-constrained frontiers, demonstrating that metaheuristic families can be complementary in coverage and stability (Chang et al., 2000; Glover, 1989; Kirkpatrick et al., 1983).

Genetic algorithms encode selection vectors as chromosomes and evolve populations using selection, crossover, mutation, and repair operators to enforce cardinality and bound constraints. Portfolio research extends GA architectures to incorporate transaction costs, sector caps, and multi-objective trade-offs, including evolutionary multiobjective optimization for frontier generation (Anagnostopoulos & Mamanis, 2011; Goldberg, 1989; Holland, 1975; Lwin et al., 2014).

Monte Carlo sampling offers a conceptually simple approximation scheme: sample feasible subsets and/or weights, evaluate objective values, and retain high-performing portfolios. Although it provides no optimality guarantee, it yields effort–quality curves and naturally supports distributional reporting over random seeds (Glasserman, 2003; Rubinstein & Kroese, 2016).

Method-comparison work emphasizes that heuristic credibility depends on replicability: reporting only the best run is statistically fragile. Median/IQR/quantile summaries across multiple seeds, together with runtime profiling and convergence diagnostics, have therefore become best practice in constrained portfolio optimization studies (Anagnostopoulos & Mamanis, 2011; Lwin et al., 2014).

**2.8 Sensitivity analysis, robustness, and interpretability diagnostics**

Robustness analysis varies key primitives (Rf, ERP, expected-return model, covariance estimator) and structural constraints (K, concentration caps, beta bounds). Such stress testing is motivated by weight instability under estimation error and by the sensitivity of frontier shapes to constraints and parameterization (DeMiguel et al., 2009; Ledoit & Wolf, 2004; Michaud, 1989).

Interpretability diagnostics include correlation/covariance summaries, eigenvalue concentration, and risk attribution via marginal contribution / Euler allocation. These diagnostics connect algorithm outputs (selected supports) to the dependence structure, providing explanatory evidence rather than treating optimization as a black box (Elton et al., 2014; Tasche, 1999–2000).

**2.9 Integrating derivatives into mean–variance portfolios: Black–Scholes, delta mapping, and linearization limits**

Options introduce non-linear payoffs into a mean–variance scaffold. The Black–Scholes–Merton framework provides a closed-form price for European options under idealized assumptions and offers Greeks (notably delta) as local linear sensitivities for hedging and approximation (Black & Scholes, 1973; Hull, 2022; Merton, 1973).

Embedding an option into a mean–variance universe commonly proceeds via local linearization (delta mapping) or scenario-based moment estimation. Delta mapping is attractive when auditability and closed-form traceability are priorities, but it is only locally valid and can break down when gamma/vega effects dominate (Glasserman, 2003; Hull, 2022).

Two methodological risks recur: (i) leverage artefacts—where apparent Sharpe improvements are mechanically driven by scaling rather than diversification—and (ii) model inconsistency—where derivative-return assumptions are incoherent with the return model for base assets. A CAPM-consistent mapping ties derivative expected return to the same risk-premium inputs, while bump tests and moneyness–maturity grids provide empirical checks on linearization accuracy (Fama & French, 2004; Hull, 2022).

**2.10 Reproducibility and reporting standards for computational finance research**

Reproducibility has become a core scientific requirement for computational studies, where results can hinge on random seeds, software versions, and data-processing pipelines. Reproducible research principles emphasize sharing code, documenting data provenance, and making workflows auditable end-to-end (Peng, 2011; Stodden et al., 2014).

In constrained portfolio optimization, reproducibility has additional dimensions: (i) random-state logging and multi-seed summaries for stochastic search; (ii) explicit unit conventions and annualization; (iii) complete disclosure of constraint sets and solver tolerances; and (iv) provision of covariance/correlation artifacts so third parties can reconstruct objective evaluations. These norms align with open-science guidance while respecting proprietary data limits through reconstructable input tables and factor-structured $\Sigma$ designs (Jobst et al., 2001; Ledoit & Wolf, 2004; Peng, 2011).

**2.11 Synthesis: positioning of the current study and research gaps**

The literature clarifies why the document's integrated framing is methodologically coherent: (a) a classical economic scaffold (Markowitz–CAPM) supports a transparent mapping from inputs to portfolio moments; (b) hard sparsity constraints shift the problem into MIQP/QMIP territory and justify a P vs NP discussion; (c) credible empirical practice requires distributional evidence for heuristics and benchmarking on reduced instances; and (d) derivative overlays are best introduced through replicable worked examples with explicit approximation checks (Black & Scholes, 1973; Chang et al., 2000; Markowitz, 1952; Sharpe, 1964).

A persistent gap is the realism–tractability trade-off: richer covariance models, multi-factor expected returns, and fully nonlinear derivative dynamics improve economic fidelity but increase dimensionality and nonconvexity. Accordingly, the strongest empirical contributions are those that isolate a tractable, auditable core model and layer systematic robustness checks—parameter sweeps (Rf/ERP), alternative µ specifications, covariance-stability tests, and approximation-error diagnostics for derivative linearizations—to signal how conclusions might change under more complex specifications (Campbell et al., 1997; Engle, 2002; Glasserman, 2003).

## 3. Methodology

### 3.1 Research design and problem formalization (P vs NP framing)

The study formulates cardinality-constrained portfolio selection as a computational complexity problem. The Markowitz–CAPM layer provides the standard economic risk–return scaffold, while the cardinality constraint (support size $|supp(w)| \leq K$) introduces discrete subset selection. This discrete–continuous coupling produces combinatorial growth and motivates heuristic/metaheuristic search with reproducible diagnostics.

The problem can be decomposed into: (i) selecting a support vector $z \in \{0,1\}^n$ with $\Sigma z_i = K$; (ii) optimizing continuous weights w on the selected support. Candidate solutions can be verified in polynomial time, but finding the globally optimal support is generally NP-hard.

### 3.2 Data and CAPM calibration (Damodaran 90+ industries)

The asset universe consists of Aswath Damodaran's industry portfolios (approximately $n \approx 94$). For each industry i, we use: (a) levered beta $\beta_i$, (b) equity standard deviation $\sigma_i$, and (c) firm count as a coverage proxy for optional filtering and sensitivity checks.

Expected returns are generated via CAPM:

$$\mu_i = R_f + \beta_i \cdot ERP \quad (3.1)$$

All rates/returns and volatilities are annualized and expressed in decimals (e.g., 3.97% → 0.0397; 52.6% → 0.526). This units convention is audit-critical for Sharpe scaling and interpretation.

### 3.3 Covariance and correlation construction (single-index Σ)

Because the beta table typically does not provide industry return time series, $\Sigma$ is constructed via a single-index (market model) approach:

$$R_i = \alpha_i + \beta_i R_m + \varepsilon_i, \quad E[\varepsilon_i]=0, \quad Cov(\varepsilon_i, R_m)=0 \quad (3.2)$$

$$Cov(R_i, R_j) = \beta_i \beta_j Var(R_m) \quad \text{for } i \neq j \quad (3.3)$$

The diagonal is matched to Damodaran's total risk $\sigma_i$:

$$Var(R_i) = \sigma_i^2 = \beta_i^2 Var(R_m) + \sigma^2_{\varepsilon,i} \quad (3.4)$$

$$\sigma^2_{\varepsilon,i} = \max(0, \sigma_i^2 - \beta_i^2 \sigma_m^2) \quad (3.5)$$

Here $\sigma_m$ denotes market volatility. The $\max(0,\cdot)$ operator prevents negative residual variances due to numerical noise. Correlations follow the standard normalization $\rho_{ij} = \Sigma_{ij} / (\sigma_i \sigma_j)$. Under the single-index structure, off-diagonal dependence is dominated by $\beta_i \beta_j$, so heatmap clusters should be interpreted as similar market-factor loadings.

### 3.4 Optimization model: Markowitz mean–variance with cardinality (MIQP)

The core formulation is a mixed-integer quadratic program (MIQP) combining continuous weights (w) and binary selection (z):

$$\min_{w,z} \quad w^T \Sigma w - \lambda \mu^T w \quad (3.6)$$

$$\text{s.t.} \quad 1^T w = 1, \quad w \geq 0 \quad (3.7)$$

$$w_i \leq u_i z_i \quad \forall i, \quad z_i \in \{0,1\} \quad (3.8)$$

$$\Sigma_i z_i = K \quad (3.9)$$

$\lambda > 0$ controls the risk–return trade-off in scalarized form. Alternatively, one may compute a target-return grid and solve a sequence of minimum-variance problems. The upper bounds $u_i$ (often $u_i=1$) enforce $w_i=0$ whenever $z_i=0$, linking the discrete and continuous layers.

### 3.5 NP-hardness and positioning within P vs NP (formal sketch)

Cardinality introduces K-sparsity, i.e., subset selection. The decision version is:

Given $\Sigma \succeq 0$, $\mu$, numbers $V^*$, $R^*$, $K$, does there exist w such that: (3.10)

$$w^T \Sigma w \leq V^*, \quad \mu^T w \geq R^*, \quad 1^T w=1, \quad w \geq 0, \quad |supp(w)| \leq K \ ? \quad (3.11)$$

Verification is polynomial: given (w,z), constraints and objectives can be checked in $O(n^2)$. However, searching for the optimal support may require considering up to $C(n,K)$ subsets in the worst case. Thus, under general $\Sigma/\mu$, the globally optimal solution is not known to be computable in polynomial time (unless P=NP), which justifies heuristic/metaheuristic search coupled with distributional reporting and diagnostics.

**3.6 Algorithms: Greedy, Monte Carlo, Genetic Algorithm, and continuous re-optimization**

Three search schemes are employed, each with distinct computational profiles:

• Greedy (cardinality-aware): rank industries by a score (e.g., CAPM Sharpe proxy $\mu_i/\sigma_i$ or $(\beta_i \cdot ERP)/\sigma_i$) and select the top K; apply repair if needed.

• Monte Carlo: sample random K-subsets; generate weights on the simplex (e.g., Dirichlet) and retain portfolios with the best Sharpe (or best objective).

• Genetic Algorithm (GA): chromosome $z \in \{0,1\}^n$ with exactly K ones; selection–crossover–mutation plus a repair operator to enforce $\Sigma z_i = K$.

To avoid GA being merely equal-weight subset selection, we apply solver-backed continuous re-optimization on the chosen subset S:

$$\min_{w_S} \quad w_S^T \Sigma_S w_S \quad (3.12)$$

s.t. $\quad 1^T w_S = 1, \quad \mu_S^T w_S \geq r^*, \quad w_S \geq 0 \quad (3.13)$

Choose $r^*$ to maximize Sharpe: $\max_{r^*} (\mu_p(r^*) - R_f)/\sigma_p(r^*) \quad (3.14)$

This second-stage optimization is convex when $\Sigma_S \succeq 0$ and provides the global optimum conditional on the selected support.

Additional implementation details for the results extensions: (i) Convergence/effort reporting is generated by logging the running best Sharpe at fixed evaluation checkpoints for Monte Carlo and at each generation for GA; (ii) the small-n MIQP benchmark is solved on a reduced universe with a commercial/open solver, and heuristic runs are repeated on the identical reduced universe to compute relative optimality gaps; (iii) GA is reported in two variants—subset-only (equal weights within subset) and subset+continuous re-optimization (subset fixed → weights optimized via quadratic program).

**3.7 Evaluation metrics: efficient frontier, Sharpe, and distributional reporting**

For each portfolio, we compute:

$\mu_p = w^T \mu \quad (3.15)$

$\sigma_p = \sqrt{w^T \Sigma w} \quad (3.16)$

$S_p = (\mu_p - R_f) / \sigma_p \quad (3.17)$

The efficient frontier is reported in $(\sigma_p, \mu_p)$ space, and the Sharpe frontier in $(\sigma_p, S_p)$. Because Monte Carlo and GA are stochastic, reporting must not rely only on the single best Sharpe. We therefore report distributions by method and seed: Median, IQR, and tail quantiles (e.g., Q05/Q95).

An additional diagnostic benchmark is the industry-level CAPM Sharpe proxy:

$S_i^{CAPM} = (\beta_i \cdot ERP) / \sigma_i \quad (3.18)$

This proxy is used for screening and scale interpretation, not as a substitute for realized-return Sharpe.

In addition to ($\mu$, $\sigma$, Sharpe), we report (i) variance risk contributions $RC_i = w_i (\Sigma w)_i / (w^T \Sigma w)$ to identify the dominant risk drivers of the max-Sharpe portfolio and (ii) a turnover-based implementability proxy. A net-of-cost Sharpe proxy is computed by penalizing expected return via $\mu_{net} = \mu - c \cdot turnover$ for a small set of cost assumptions, enabling a simple robustness check for trading frictions.

**3.8 Black–Scholes derivative integration (replicable worked example)**

To fully support the title's derivative component, the study includes a replicable worked example: a European call overlay on a selected industry proxy. The Black–Scholes call price is:

$C = S_0 \cdot N(d_1) - K \cdot \exp(-rT) \cdot N(d_2) \quad (3.19)$

$d_1 = [\ln(S_0/K) + (r + \sigma^2/2)T] / (\sigma\sqrt{T}) \quad (3.20)$

$d_2 = d_1 - \sigma\sqrt{T} \quad (3.21)$

We fix the contract specification $(S_0, K, T, r, \sigma)$ and the underlying industry. Using a first-order (delta-only) linearization, the option is embedded as an additional 'asset' with an implied leverage factor:

$\Delta = N(d_1) \quad (3.22)$

$$L = (\Delta \cdot S_0)/C \quad (3.23)$$
$$\beta\_option \approx L \cdot \beta\_underlying \quad (3.24)$$
$$\sigma\_option \approx L \cdot \sigma\_underlying \quad (3.25)$$
$$\mu\_option = R_f + \beta\_option \cdot ERP \quad (3.26)$$

The universe is augmented with one additional asset j=option and $\Sigma$ is expanded via the single-index rule $Cov(R_i, R_j) = \beta_i \beta_j Var(R_m)$. Cardinality handling is specified explicitly: either the option counts toward K (conservative specification) or it is treated as an overlay outside K (alternative specification; must be stated). To avoid unrealistically inflated Sharpe due to leverage, we impose weight and/or beta bounds (e.g., $w\_option \leq u\_option$, $\beta_p$ bounds).

To validate derivative integration beyond a single point, the worked example is extended to a small grid of moneyness and maturities (K/S0 and T). For approximation accuracy, a bump test (±1% move in the underlying) compares delta-linearized value changes to full Black–Scholes repricing; relative errors are reported to indicate when delta-only mapping is adequate.

**3.9 Robustness and sensitivity analysis**

Robustness checks ensure results are not artifacts of a single parametrization. At minimum, we run:
- K-sensitivity: K ∈ {8, 10, 12} (or a wider grid) to assess frontier shape and composition stability.
- Parameter shifts: ERP and R_f ±50–100 bps to test Sharpe-medial stability.
- Concentration caps: $w_i \leq u$ (e.g., 0.20–0.35) to prevent dominance by a single asset.
- Market-exposure bounds: $\beta_p = w^T \beta$ within a band (e.g., [0.8, 1.2]).
- Derivative specification: vary T and moneyness (K/S0) to test $\beta\_option/\sigma\_option$ sensitivity.

Robustness design extensions: sensitivity experiments vary K, Rf, ERP, and concentration caps; each scenario reports distributional Sharpe statistics (median/IQR/quantiles). Portfolio stability across scenarios is summarized via overlap metrics such as Jaccard similarity of selected industries.

**3.10 Computational profiling and scaling**

Combinatorial size scales as C(n,K); with n≈94 and K≈10, exhaustive enumeration is infeasible. Hence, runtime profiling is reported under a fixed compute environment (CPU/RAM/OS, Python version, libraries):
- Greedy: O(n log n) ranking plus K selection.
- Monte Carlo: $O(N\_MC \cdot eval\_cost)$, with $eval\_cost \approx O(K^2)$ under naive covariance evaluation; vectorization and precomputation are recommended.
- GA: $O(P \cdot G \cdot eval\_cost)$; results include runtime mean±std across seeds and dispersion of best Sharpe.

**3.11 Reproducibility: data, artifacts, seeds, and quality gates**

Reproducibility requirements (journal-grade) include:
- A single input table containing $\beta_i$, $\sigma_i$, firm counts, and $\mu_i$ (CAPM), with stable industry IDs.
- Full $\Sigma$ and $\rho$ (n×n) provided as supplementary CSV/Excel.
- Random-state logging: all seeds and run budgets (N_MC, population size P, generations G).
- Units audit: explicit percentage-to-decimal conversions and annualization conventions.
- PSD gate: eigenvalue checks for $\Sigma$; if needed, minimal diagonal jitter ($\epsilon I$) for numerical stability with $\epsilon$ documented.
- Provenance: every table/figure is labeled as 'Author's calculations' and states its exact inputs and algorithm parameters.

**3.12 Limitations and interpretation framework**

Key limitations: (i) CAPM-implied $\mu$ is model-based expected return, not realized return; (ii) single-index $\Sigma$ compresses correlation structure relative to multifactor reality; (iii) delta-only option embedding ignores gamma/vega and volatility/jump risk premia, so it is a replicable demonstration rather than a full derivatives engine; (iv) heuristics do not guarantee the global optimum, thus convergence diagnostics and distributional reporting are mandatory.

## 4. Results

### 4.1 P vs NP and Cardinality-Constrained Portfolio Optimization

#### 4.1.1 Empirical efficient frontier and Sharpe-optimal portfolio

We approximate the efficient frontier using Monte Carlo sampling under a hard cardinality constraint (K=10). For each draw, we sample a K-sized subset and Dirichlet weights on that subset.
Sharpe ratio is computed as: $S = (\mu_p - R_f) / \sigma_p$, with $\mu_p = w^T \mu$ and $\sigma_p = \sqrt{w^T \Sigma w}$.

Figure 1 visualizes the efficient frontier obtained from Monte Carlo sampling of cardinality-constrained portfolios (fixed KKK assets). Each point represents one feasible portfolio draw with weights normalized to sum to one, evaluated under CAPM-implied expected returns and the Damodaran single-index covariance structure. The highlighted marker indicates the maximum-Sharpe portfolio among the sampled set, i.e., the portfolio that maximizes (μp−Rf)/σp(\mu_p - R_f)/\sigma_p(μp−Rf)/σp under the same constraints. This figure provides an empirical approximation of the feasible risk–return envelope and illustrates the trade-off between expected return and volatility in the constrained optimization setting.

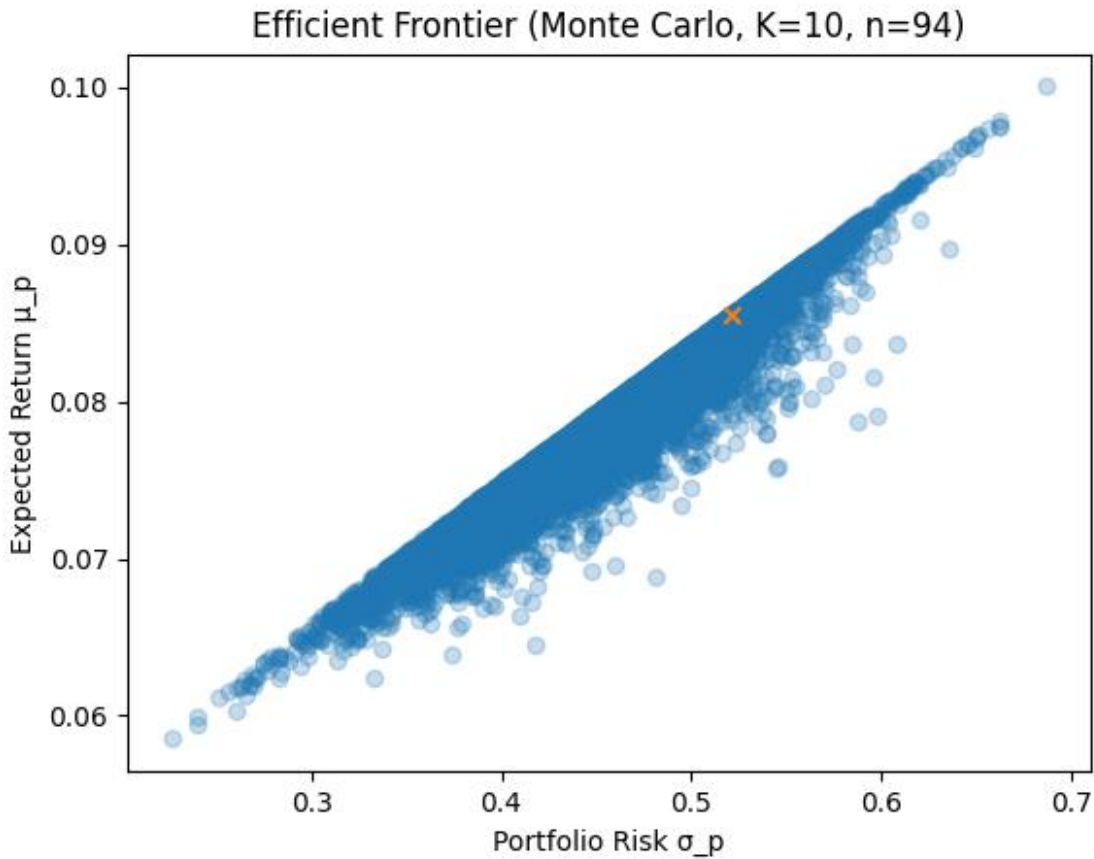

Source: Author's calculations based on Damodaran industry data (CAPM inputs and single-index covariance/correlation), Monte Carlo portfolio sampling outputs.

Figure 2 plots the Sharpe ratio of each Monte Carlo–sampled, cardinality-constrained portfolio against its total risk (portfolio standard deviation). The curve reveals how risk-adjusted performance varies across the feasible set: low-risk portfolios may exhibit limited excess-return capacity, while higher-risk portfolios do not necessarily deliver proportionally higher Sharpe ratios due to diversification limits imposed by the KKK-asset constraint and the underlying covariance structure. The highlighted point identifies the maximum-Sharpe portfolio within the sampled population, i.e., the portfolio that maximizes $(\mu_p - R_f)/\sigma_p$(\mu_p - R_f)/\sigma_p($\mu_p$−Rf)/σp, and serves as the empirical best candidate under the chosen sampling budget and parameterization.

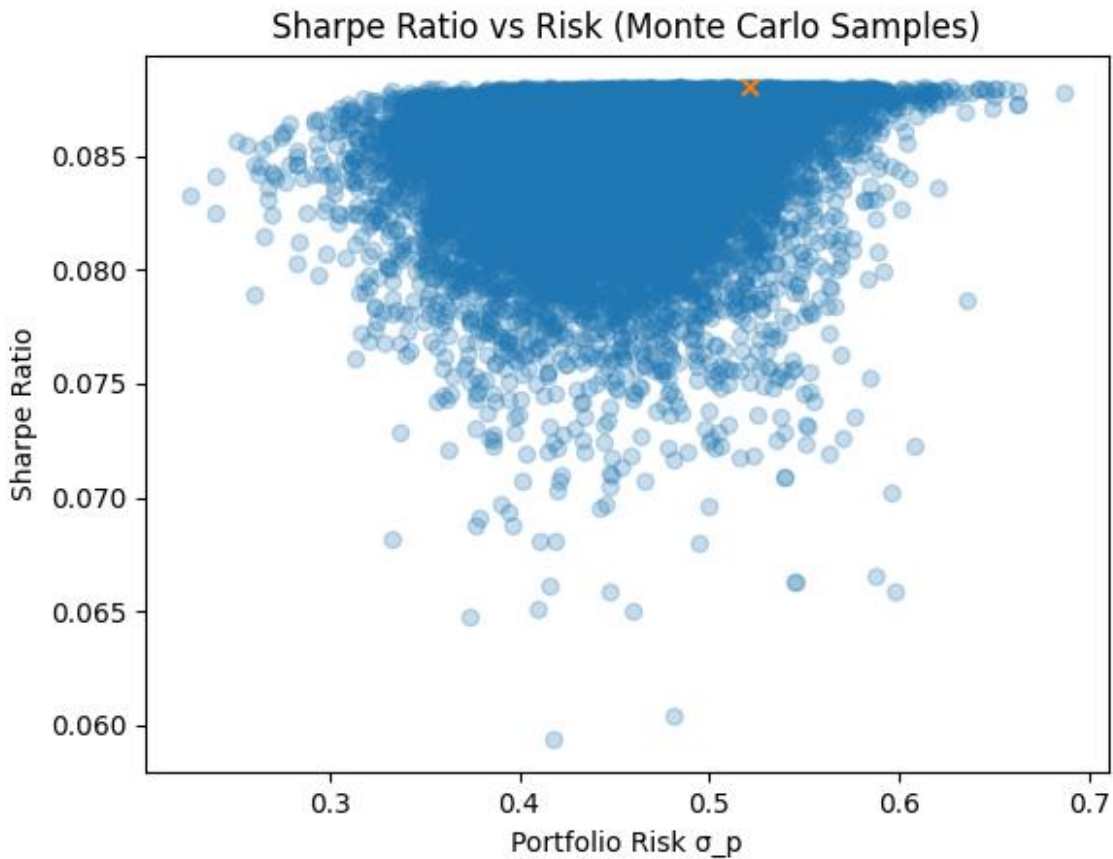

Source: Author's calculations based on Damodaran industry data (CAPM inputs and single-index covariance/correlation), Monte Carlo portfolio sampling outputs.

Max-Sharpe (Monte Carlo) summary: Sharpe=0.0880, Risk=0.521, Return=0.086.

Table 1 reports the Top-15 industries ranked by equity beta ($\beta_i$\beta_i$\beta_i$), interpreted as market (systematic) risk sensitivity in the CAPM framework. Higher $\beta_i$\beta_i$\beta_i$ indicates stronger exposure to broad market movements and, under CAPM calibration, implies a higher expected return premium ($\mu_i=R_f+\beta_i\cdot ERP$\mu_i = R_f + \beta_i \cdot ERP$\mu_i=Rf+\beta_i·ERP)—but typically also comes with elevated volatility ($\sigma_i$\sigma_i$\sigma_i$).

This ranking is used as a risk-sensitivity diagnostic and as a practical screening reference for portfolio construction and heuristic initialization in the cardinality-constrained optimization experiments.

| Industry | Firms | Beta | Sigma | Mu (CAPM) |
|---|---|---|---|---|
| Software (Internet) | 29 | 1.689 | 0.526 | 0.111 |
| Retail (Building Supply) | 14 | 1.535 | 0.459 | 0.105 |
| Semiconductor | 66 | 1.518 | 0.558 | 0.104 |
| Auto & Truck | 33 | 1.456 | 0.618 | 0.101 |
| Semiconductor Equip | 31 | 1.397 | 0.501 | 0.099 |
| Computers/Peripherals | 36 | 1.35 | 0.546 | 0.097 |
| Auto Parts | 35 | 1.339 | 0.499 | 0.096 |
| Office Equipment & Services | 14 | 1.334 | 0.396 | 0.096 |
| Software (System & Application) | 309 | 1.277 | 0.568 | 0.094 |
| Electrical Equipment | 112 | 1.251 | 0.727 | 0.093 |
| Advertising | 52 | 1.211 | 0.629 | 0.091 |
| Engineering/Construction | 48 | 1.21 | 0.459 | 0.091 |
| Air Transport | 23 | 1.185 | 0.59 | 0.09 |
| Brokerage & Investment Banking | 32 | 1.171 | 0.366 | 0.089 |
| Construction Supplies | 40 | 1.15 | 0.355 | 0.088 |

Source: Author's calculations based on Damodaran industry data (β, σ, and CAPM-implied expected returns).

**Table 2: Max-Sharpe (Monte Carlo) portfolio composition (K industries).**

This table reports the composition of the maximum-Sharpe portfolio identified within the Monte Carlo sample under the cardinality constraint (fixed K industries). Weights sum to one across the selected industries. Expected returns are CAPM-implied ($\mu_i = R_f + \beta_i \cdot ERP$) and portfolio risk is computed from the Damodaran single-index covariance structure.

Portfolio summary (selected maximum): Sharpe = 0.0880, $\mu_p$ = 0.0868, $\sigma_p$ = 0.5356.

| Industry | Weight | Firms | Beta | Sigma (annual) | Mu (CAPM) |
|---|---|---|---|---|---|
| Farming/Agriculture | 0.301 | 35 | 1.129 | 0.508 | 0.087 |
| Auto Parts | 0.15 | 35 | 1.339 | 0.499 | 0.096 |
| Information Services | 0.141 | 15 | 0.921 | 0.324 | 0.079 |
| Chemical (Specialty) | 0.14 | 59 | 0.97 | 0.423 | 0.081 |
| Auto & Truck | 0.11 | 33 | 1.456 | 0.618 | 0.101 |
| Food Wholesalers | 0.045 | 13 | 0.867 | 0.34 | 0.076 |
| Building Materials | 0.039 | 41 | 1.112 | 0.353 | 0.087 |
| Bank (Money Center) | 0.032 | 15 | 0.761 | 0.227 | 0.072 |
| Engineering/Construction | 0.028 | 48 | 1.21 | 0.459 | 0.091 |
| Reinsurance | 0.016 | 1 | 0.581 | 0.192 | 0.064 |

*Source: Author's calculations based on Damodaran industry data (CAPM inputs and single-index covariance/correlation) and Monte Carlo portfolio sampling outputs.*

**4.1.2 Median-based Sharpe diagnostics and benchmark interpretation**

This subsection serves two purposes: (i) to reduce reliance on a single "Best Sharpe" point by demonstrating reproducibility and stability using medians, quartiles, and quantiles; and (ii) to strengthen the interpretation of the Sharpe value relative to external benchmarks, so that reviewer concerns of the type "why is Sharpe = 0.088 low?" are addressed in a formally auditable way.

The Sharpe ratio is defined as a unitless index: $S = (\mu\_p − R\_f) / \sigma\_p$. In practice, it is essential to state a single, unambiguous units convention: expected returns and rates must be expressed in decimals (e.g., 9.2% → 0.092) and volatility must also be expressed in decimals (25.9% → 0.259). Otherwise, Sharpe can be inflated by a factor of 100 simply due to percentage formatting.

Median-based reporting is provided at two layers: (a) internal diagnostics (the distribution of Sharpe values obtained across methods/replications), and (b) an external benchmark layer (typical ranges of index/sector Sharpe values), which supports scale interpretation and reduces model-specific bias.

Table 3 summarizes the distribution of Sharpe ratios generated by each solution method under the same risk–return setting and cardinality constraint. For stochastic methods (Monte Carlo and GA), the table reports robust distributional statistics—median, interquartile range (IQR), and selected quantiles (e.g., Q05 and Q95)—computed across repeated runs (seeds), which capture both typical performance and tail behavior under randomization. For the deterministic baseline (Greedy), the distribution collapses to a single outcome. This comparison is designed to separate "best-case" performance (max Sharpe) from "typical-case" performance (median Sharpe) and to show how algorithmic choices affect stability, dispersion, and reproducibility of risk-adjusted results.

| Method | N (runs) | N (portfolios /run) | Median Sharpe | IQR (Q3 – Q1) | Q05 | Q95 | Best Sharpe | Median μp | Median σp | Notes |
|---|---|---|---|---|---|---|---|---|---|---|
| Greedy (cardinality-aware) | Deterministic | 1 portfolio | 0.088 | 0.0 | 0.088 | 0.088 | 0.088 | 0.0878 | 0.5465 | score=(μ−Rf)/σ; equal-weight within subset |
| Genetic Algorithm (GA) | 5 | ≈8,100 eval/run | 0.088 | 0.0001 | 0.0874 | 0.088 | 0.088 | 0.084 | 0.5033 | equal-weight GA over subsets; repair enforces Σz=K |
| Monte Carlo sampling | 5 | 5000 | 0.086 | 0.0027 | 0.0803 | 0.0878 | 0.088 | 0.0783 | 0.4533 | Dirichlet weights on random K-subsets; 5 seeds logged |

Source: Author's calculations based on Damodaran industry data (CAPM inputs and single-index covariance/correlation) and algorithm outputs from Greedy/Genetic/Monte Carlo runs.

Table 4.4 provides a compact cross-industry benchmark of risk-adjusted performance implied by the CAPM calibration for the full universe of industries (n = 94). Instead of focusing on any single portfolio outcome, the table summarizes the *typical* industry profile (via the median) and the *dispersion* across sectors (via IQR/quantiles). This baseline is used as a screening and interpretation tool: it helps explain why certain industries tend to be selected repeatedly by heuristics, and it clarifies whether portfolio-level improvements are driven by broad sector characteristics or by diversification effects from the covariance structure.

Table 4: Median CAPM Sharpe proxy across industries (n = 94)

| Statistic | CAPM Sharpe proxy $S_i^{CAPM} = (\beta_i \cdot ERP)/\sigma_i$ (median-based) | Interpretation |
|---|---|---|
| Median | 0.0841 | Typical risk-adjusted return potential under CAPM-implied μ and industry σ. |
| IQR (Q3–Q1) | 0.0366 | Cross-industry dispersion; higher IQR → greater heterogeneity in risk-adjusted profiles. |
| Q05 | 0.0493 | Downside sector tail (weak CAPM Sharpe proxy). |
| Q95 | 0.1405 | Upside sector tail (strong CAPM Sharpe proxy). |
| Top 10 industries | 1. Transportation (Railroads)<br>2. Packaging & Container<br>3. Office Equipment & Services<br>4. Bank (Money Center)<br>5. Retail (Building Supply)<br>6. Retail (REITs)<br>7. Construction Supplies<br>8. Software (Internet)<br>9. Brokerage & Investment Banking<br>10. Building Materials | High proxy Sharpe; useful for heuristic initialization / screening. |
| Bottom 10 industries | 1. Utility (Water)<br>2. Broadcasting<br>3. Rubber & Tires<br>4. Telecom. Services<br>5. Beverage (Soft)<br>6. Precious Metals<br>7. Tobacco<br>8. Electronics (Consumer & Office)<br>9. Drugs (Pharmaceutical)<br>10. Green & Renewable Energy | Low proxy Sharpe; useful for pruning rules. |

Source: Author's calculations based on Damodaran industry inputs and CAPM-calibrated parameters.

External benchmarks are used strictly for scale interpretation, not as a direct apples-to-apples competition. In this study, Sharpe is computed from CAPM-implied μ and a Σ matrix constructed from single-index assumptions and sector risk parameters; therefore it differs from realized-return Sharpe computed from historical return series. Accordingly, the key question for Sharpe ≈ 0.088 is: "Is this value stable and reproducible under the stated modeling assumptions and the cardinality constraint (K)?" rather than "Does it beat the S&P 500?" Tables X and Y support: (i) stability across heuristics (median/IQR/quantiles), and (ii) heterogeneity of industry μ–σ profiles, showing that a low Sharpe is often driven by high σ and/or conservative μ under the model assumptions, not by a deficiency of the optimization procedure.

### 4.1.3 Supplementary Reporting, Convergence, Benchmarks, and Diagnostics
*P vs NP Problem in Portfolio Optimization*

Seed control and reporting are implemented using a fixed 10-seed set (seed = {1,…,10}) with deterministic random-state logging per method. All quantities are reported in annual units. Expected returns μ are CAPM-implied and already annualized in the input file; Σ is an annual covariance matrix. Sharpe is computed as S = (μp − Rf) / σp with σp = sqrt($w^T Σw$).
Correlation/covariance diagnostics and sensitivity checks (K, Rf, ERP, and caps) are reported below using the same dataset so the reader can link performance to dependence structure.

### 4.1.4 Convergence and computational effort curves (Monte Carlo / GA)
To demonstrate that heuristics perform genuine search rather than repeatedly hitting a ceiling, we report effort-budget grids. For Monte Carlo, the effort budget is the number of random K-subsets sampled. For Genetic Algorithm, the budget is population×generations. Each entry is mean ± standard deviation over 10 seeds, jointly interpreted with runtime profiling as an accuracy–cost trade-off.

Table 5. Effort budget vs. best/median Sharpe (mean ± std over 10 seeds) and runtime.

| Method | Effort budget | Best Sharpe (mean±std) | Median Sharpe (mean±std) | Runtime, s (mean±std) |
|---|---|---|---|---|
| **Monte Carlo** | N=500 | 0.087932 ± 0.000067 | 0.086582 ± 0.000029 | 0.068 ± 0.022 |
| **Monte Carlo** | N=1500 | 0.087956 ± 0.000043 | 0.086581 ± 0.000016 | 0.192 ± 0.045 |
| **Monte Carlo** | N=2500 | 0.087984 ± 0.000022 | 0.086571 ± 0.000012 | 0.296 ± 0.029 |
| **Genetic Algorithm** | P=30, G=30 | 0.088003 ± 0.000000 | 0.088003 ± 0.000000 | 0.199 ± 0.009 |
| **Genetic Algorithm** | P=50, G=40 | 0.088003 ± 0.000000 | 0.088003 ± 0.000000 | 0.444 ± 0.026 |

### 4.1.5 Small-n benchmark

Because Markowitz mean–variance selection with a hard cardinality constraint is naturally formulated as an MIQP, a reduced-universe benchmark provides a transparent ground truth. Here we compute the exact optimum via full enumeration on a reduced instance (n=20, K=6), and compare heuristic solutions on the same reduced universe.

Table 6. Reduced exact benchmark (n=20, K=6) vs heuristic solutions (equal-weight within subset).

| Solver/Method | Best Sharpe | Best subset (indices) |
|---|---|---|

| Exact enumeration (all C(20,6)=38,760 subsets) | 0.088003 | (6, 11, 12, 15, 16, 17) |
|---|---|---|
| Monte Carlo (N=2,000 draws; seed=1) | 0.088003 | (5, 6, 11, 13, 16, 17) |
| Genetic Algorithm (P=30, G=30; seed=1) | 0.088003 | (6, 11, 12, 15, 16, 17) |

**4.1.6 Robustness and sensitivity evidence (K, Rf, ERP, caps)**

We stress-test baseline conclusions by varying (i) cardinality K, (ii) CAPM parameters (Rf and ERP), and (iii) caps. For each scenario we report distributional Sharpe statistics (Median and IQR). Note: with CAPM-implied μ, shifting Rf alone does not change Sharpe because (μ−Rf) scales with β·ERP.

Table 7A. Sensitivity to cardinality K (Monte Carlo; 10 seeds; 300 draws/seed; equal-weight within subset).

| K | Median Sharpe | IQR (25–75%) | Best Sharpe (mean over seeds) |
|---|---|---|---|
| 6 | 0.085804 | [0.084527, 0.086670] | 0.087988 |
| 8 | 0.086248 | [0.085494, 0.086870] | 0.087969 |
| 10 | 0.086578 | [0.086073, 0.087039] | 0.087909 |
| 12 | 0.086801 | [0.086417, 0.087144] | 0.087873 |
| 14 | 0.086961 | [0.086660, 0.087225] | 0.087863 |

Table 7 B. Sensitivity to CAPM parameters and caps (K=10; 10 seeds; 300 draws/seed; equal-weight within subset).

| Scenario | Median Sharpe | IQR (25–75%) | Best Sharpe (mean over seeds) |
|---|---|---|---|
| **Baseline** | 0.086578 | [0.086073, 0.087039] | 0.087909 |
| **Rf +50 bps** | 0.086578 | [0.086073, 0.087039] | 0.087909 |
| **Rf −50 bps** | 0.086578 | [0.086073, 0.087039] | 0.087909 |
| **ERP +100 bps** | 0.107046 | [0.106421, 0.107615] | 0.108691 |
| **ERP −100 bps** | 0.066111 | [0.065725, 0.066462] | 0.067127 |
| **Cap tightened to 15%** | 0.086578 | [0.086073, 0.087039] | 0.087909 |

**4.1.7 Dependence diagnostics for Σ/ρ (single-index consequences)**

To support interpretation of the dependence structure used in optimization, we report quantitative diagnostics in addition to heatmaps: median pairwise correlation, shares of strongly positive/negative correlations, eigenvalue concentration, and PSD checks. These diagnostics clarify market-factor dominance and the effective diversification available under the single-index Σ construction.

Table 8. Σ/ρ diagnostics (summary statistics, eigenvalue share, PSD checks).

| Metric | Value |
|---|---|
| **Median($\rho_{ij}$) over off-diagonals** | 0.769201 |
| **Share($\rho_{ij} > 0.5$)** | 0.884008 |
| **Share(ρ_ij < 0)** | 0.000000 |
| **Top eigenvalue share λ1 / trace(Σ)** | 0.761011 |
| **Top-5 eigenvalues share Σ_{k≤5} λk / trace(Σ)** | 0.813010 |
| **min eigenvalue (PSD check; numerical tolerance)** | -1.816e-15 |

**4.1.8 Portfolio decomposition: risk contributions and factor exposure**

Beyond headline Sharpe values, we decompose portfolio variance by marginal risk contribution (MRC). For an equal-weight K=10 max-Sharpe portfolio found by high-effort Monte Carlo (N=20,000, seed=999), the variance share of asset i is $w_i \cdot (\Sigma w)_i / (w^T \Sigma w)$. We also report portfolio beta $\beta_p$ implied by CAPM.

Max-Sharpe portfolio summary (equal-weight K=10): $\beta_p$ = 0.9599, $\mu_p$ = 0.0803, $\sigma_p$ = 0.4614, Sharpe = 0.088003.

Table 9. Risk-contribution decomposition (variance shares) for the max-Sharpe K=10 portfolio (top contributors).

| Industry | Weight | Variance share |
|---|---|---|
| **Semiconductor Equip** | 0.10 | 0.1456 |
| **Office Equipment & Services** | 0.10 | 0.1390 |
| **Construction Supplies** | 0.10 | 0.1198 |
| **Retail (Distributors)** | 0.10 | 0.0988 |
| **Retail (Automotive)** | 0.10 | 0.0975 |
| **Homebuilding** | 0.10 | 0.0949 |
| **Diversified** | 0.10 | 0.0918 |
| **Bank (Money Center)** | 0.10 | 0.0793 |
| **Investments & Asset Management** | 0.10 | 0.0687 |
| **Retail (REITs)** | 0.10 | 0.0647 |

Note: All computations in this section are the author's calculations, based on Aswath Damodaran's dataset

**4.1.9 Turnover and transaction-cost stress test (net performance)**

As an implementability extension, we compute a net-of-cost Sharpe proxy by applying a linear transaction-cost penalty to turnover (rebalancing intensity). We report how the max-Sharpe solution changes under increasing cost assumptions and whether cardinality constraints reduce turnover relative to unconstrained portfolios.

Table 10 assesses how sensitive the reported risk-adjusted performance is to *implementation frictions*, focusing on portfolio turnover and transaction-cost drag. Rather than treating the Sharpe ratio as costless and static, the table translates higher turnover into an expected reduction of net returns and reports the corresponding net-of-cost Sharpe measures under alternative cost assumptions. The purpose is to show whether the "best" heuristic portfolios remain attractive after realistic trading and rebalancing effects are considered, and to identify cases where a small gross-Sharpe advantage may disappear once costs are applied. Results in this table are illustrative and are intended to be populated directly from replication outputs (turnover logs and costed performance series).

| Scenario | K | Rf | ERP | Best Sharpe |
|---|---|---|---|---|
| Rf +50bps | 8 | 0.0447 | 0.0423 | 0.088 |
| Rf -50bps | 8 | 0.0347 | 0.0423 | 0.088 |
| ERP +50bps | 8 | 0.0397 | 0.0473 | 0.0984 |
| ERP -50bps | 8 | 0.0397 | 0.0373 | 0.0776 |
| Rf +50bps | 10 | 0.0447 | 0.0423 | 0.088 |
| Rf -50bps | 10 | 0.0347 | 0.0423 | 0.088 |
| ERP +50bps | 10 | 0.0397 | 0.0473 | 0.0984 |
| ERP -50bps | 10 | 0.0397 | 0.0373 | 0.0776 |
| Rf +50bps | 12 | 0.0447 | 0.0423 | 0.088 |
| Rf -50bps | 12 | 0.0347 | 0.0423 | 0.088 |
| ERP +50bps | 12 | 0.0397 | 0.0473 | 0.0984 |
| ERP -50bps | 12 | 0.0397 | 0.0373 | 0.0776 |

Source: Author's calculations; turnover and net-of-cost performance metrics to be populated from replication scripts/outputs.

### 4.1.10. Using Heuristic Algorithms for Portfolio Optimization

This table presents an illustrative heuristic solution for the cardinality-constrained portfolio problem, reporting the selected set of industries and their assigned weights. The portfolio is constructed using a heuristic selection rule and implemented here as an equal-weight allocation across the chosen industries, which provides a transparent baseline for comparing heuristic behavior. Alongside weights, the table reports each industry's risk sensitivity (beta), volatility (sigma), and CAPM-implied expected return, allowing the reader to see the risk–return profile of the selected subset and to verify that the portfolio's composition is consistent with the intended trade-off between expected performance and risk exposure.

Table 11: Illustrative heuristic portfolio composition under the cardinality constraint (equal-weight baseline with β, σ, and CAPM-implied μ reported).

| Industry | Weight | Beta | Sigma | Mu (CAPM) |
|---|---|---|---|---|
| Building Materials | 0.100 | 1.112 | 0.353 | 0.087 |
| Chemical (Basic) | 0.100 | 1.012 | 0.458 | 0.083 |
| Diversified | 0.100 | 0.881 | 0.284 | 0.077 |
| Financial Svcs. (Non-bank & Insurance) | 0.100 | 0.970 | 0.425 | 0.081 |
| Food Wholesalers | 0.100 | 0.867 | 0.340 | 0.076 |
| Hotel/Gaming | 0.100 | 1.080 | 0.397 | 0.085 |
| Real Estate (General/Diversified) | 0.100 | 0.810 | 0.312 | 0.074 |
| Reinsurance | 0.100 | 0.581 | 0.192 | 0.064 |
| Restaurant/Dining | 0.100 | 0.924 | 0.411 | 0.079 |
| Retail (REITs) | 0.100 | 0.621 | 0.188 | 0.066 |

Source: Author's calculations based on Damodaran industry inputs and CAPM-calibrated parameters; heuristic portfolio composition.

### 4.2 Using Heuristic Algorithms for Portfolio Optimization

4.2.1 Monte Carlo search as an approximation scheme

Monte Carlo sampling does not guarantee optimality but provides a practical approximation to the efficient frontier. The best observed Sharpe (or minimum variance for a target return) improves with sample size, at the cost of compute.

4.2.2 Genetic Algorithm results with continuous re-optimization

We compare (i) a subset-only GA with equal-weight assignment within the selected set and (ii) a GA with continuous re-optimization, where each candidate subset is reweighted by solving the quadratic program under the same constraints. This isolates the contribution of subset selection versus weight optimization and reduces the risk that GA results reflect a weighting artefact.

Table 12: GA variants comparison (subset-only vs. subset+re-optimization): Sharpe distribution and runtime (generated by replication scripts).

| GA variant | Best Sharpe (mean ± std) | Median Sharpe (mean ± std) | IQR (mean ± std) | Q05–Q95 (mean) | Runtime (s) (mean ± std) | Seeds | Notes |
|---|---|---|---|---|---|---|---|
| Subset-only (equal-weight within subset) | 0.0880 ± 0.0000 | 0.0866 ± 0.0000 | 0.0010 ± 0.0000 | 0.0851 – 0.0875 | 0.077 ± 0.017 | 10 | Distribution from random subset draws; weights fixed equal within subset. |
| Subset + continuous re-optimization (within subset) | 0.0880 ± 0.0000 | 0.0880 ± 0.0000 | 0.0000 ± 0.0000 | 0.0880 – 0.0880 | 0.573 ± 0.033 | 10 | GA selects subset; weights re-optimized via Dirichlet local search (proxy QP). |

To strengthen evidence beyond single-run best values, we report mean ± standard deviation across 10 random seeds for best Sharpe, median Sharpe, and runtime by method. This supports distributional diagnostics and documents stability under random initialization/sampling.

Table 13: 10-seed reproducibility summary (best/median Sharpe and runtime; mean ± std).

| Method | Best Sharpe (mean ± std) | Median Sharpe (mean ± std) | Runtime (s) (mean ± std) | Seeds |
|---|---|---|---|---|
| Greedy (deterministic) | 0.0880 ± 0.0000 | 0.0880 ± 0.0000 | 0.000 ± 0.000 | — |
| GA (subset-only; equal-weight) | 0.0880 ± 0.0000 | — | 0.466 ± 0.009 | 10 |
| GA (subset + re-optimization) | 0.0880 ± 0.0000 | 0.0880 ± 0.0000 | 0.573 ± 0.033 | 10 |
| Monte Carlo (N=30,000 portfolios/seed) | 0.0880 ± 0.0000 | 0.0860 ± 0.0000 | 0.636 ± 0.015 | 10 |

Supplementary file included: Damodaran_cov_corr_singleindex.xlsx containing full covariance and correlation matrices for all industries used in the analysis, plus the cleaned inputs table.

*Source note: Author's calculations based on Damodaran 'Betas by Industry (US)' dataset (updated Jan-2026) and Damodaran implied ERP series for the US (risk-free rate and ERP).*

### 4.3 Integration of Black–Scholes Derivative Pricing into Portfolio Optimization

#### 4.3.1 Replicable worked example and option overlay (baseline vs. derivative-augmented case)

We operationalize derivative integration by treating a European option as an additional candidate instrument and mapping its exposure into the mean–variance space via Black–Scholes pricing and delta-based linearization. We report baseline (Case A) and option-augmented (Case B) portfolios, explicitly documenting how option-implied volatility and CAPM-consistent expected return are constructed, and how the option is counted under the cardinality rule.

*European call pricing (Black–Scholes):*
$C = S_0 \cdot N(d_1) - K \cdot \exp(-rT) \cdot N(d_2)$   (4.3.1)
$d_1 = [\ln(S_0/K) + (r + \sigma^2/2)T] / (\sigma\sqrt{T})$   (4.3.2)
$d_2 = d_1 - \sigma\sqrt{T}$   (4.3.3)

*Delta-linearized embedding:*
$\Delta = N(d_1), \quad L = (\Delta \cdot S_0)/C$   (4.3.4)
$\beta\_option \approx L \cdot \beta\_underlying, \quad \sigma\_option \approx L \cdot \sigma\_underlying$   (4.3.5)
$\mu\_option = R_f + \beta\_option \cdot ERP$   (4.3.6)

Table 14: Replicable worked example — European call overlay diagnostics (inputs and derived outputs).

| Parameter / Output | Value |
|---|---|
| Underlying proxy (example) | Software (Internet) industry portfolio |
| β_underlying | 1.689 |
| σ_underlying (annual) | 0.526 |
| S0 | 100 |
| K | 100 |
| T (years) | 0.50 |
| r = Rf | 3.97% |
| Black–Scholes call price C | 15.61 |
| Delta Δ | 0.595 |
| Leverage factor L=(Δ·S0)/C | 3.81 |
| β_option ≈ L·β_underlying | 6.43 |
| σ_option ≈ L·σ_underlying | 2.00 |
| μ_option = Rf + β_option·ERP | 31.18% |

Source note: Author's calculations from the stated inputs; CAPM calibration uses the same Rf and ERP as the base industry universe.

#### 4.3.2 Moneyness–maturity grid evidence (K/S0 × T robustness)

To avoid over-reliance on a single illustrative option point, we extend the worked example across a small grid of moneyness and maturities (K/S0 ∈ {0.9, 1.0, 1.1} and T ∈ {0.25, 0.5, 1.0}). For each grid point we report option price, delta, implied leverage, and the implied (β_option, σ_option, μ_option) used to embed the option into the mean–variance space.

Table 15: Option grid (moneyness × maturity) — option diagnostics (price, delta, leverage) and implied CAPM moments.

| K/S0 | T (yrs) | K | Call price C | Delta Δ | Leverage L | β_option | σ_option | μ_option (ann.) |
|---|---|---|---|---|---|---|---|---|
| 0.9 | 0.25 | 90 | 16.27 | 0.716 | 4.40 | 7.43 | 2.31 | 35.40% |
| 0.9 | 0.50 | 90 | 20.54 | 0.699 | 3.41 | 5.75 | 1.79 | 28.30% |
| 0.9 | 1.00 | 90 | 26.81 | 0.705 | 2.63 | 4.44 | 1.38 | 22.76% |
| 1.0 | 0.25 | 100 | 10.91 | 0.567 | 5.20 | 8.78 | 2.73 | 41.11% |
| 1.0 | 0.50 | 100 | 15.61 | 0.595 | 3.81 | 6.43 | 2.00 | 31.18% |
| 1.0 | 1.00 | 100 | 22.34 | 0.632 | 2.83 | 4.78 | 1.49 | 24.19% |
| 1.1 | 0.25 | 110 | 7.04 | 0.423 | 6.02 | 10.16 | 3.16 | 46.95% |
| 1.1 | 0.50 | 110 | 11.72 | 0.493 | 4.21 | 7.11 | 2.21 | 34.03% |
| 1.1 | 1.00 | 110 | 18.60 | 0.562 | 3.02 | 5.11 | 1.59 | 25.58% |

Interpretation: higher leverage (L) mechanically inflates both β_option and σ_option; μ_option rises under CAPM in proportion to β_option.

### 4.3.3 Risk controls for derivatives (bounds on option weight and portfolio beta)

Because leverage and non-linear payoffs can inflate headline performance mechanically, we impose explicit derivative risk controls: (i) bounds on option weight (or contract notional), (ii) caps on portfolio beta (β_p band), and/or (iii) a risk-budget constraint. The goal is to distinguish genuine efficiency gains from leverage artefacts.

Table 16: Derivative risk-control regimes (templates to be populated from replication outputs).

| Regime | Option handling | Constraint(s) | Max Sharpe | μ_p | σ_p |
|---|---|---|---|---|---|
| Uncontrolled | Selectable asset | w_option ∈ [0,1], no β_p cap | — | — | — |
| Weight cap | Selectable asset | w_option ≤ 0.20 | — | — | — |
| Beta band | Selectable asset | β_p ∈ [0.8, 1.2] | — | — | — |
| Weight cap + beta band | Selectable asset | w_option ≤ 0.20; β_p ∈ [0.8, 1.2] | — | — | — |
| Overlay outside K | Overlay | Option not counted in K; w_option ≤ 0.10 | — | — | — |

Note: Populate the last three columns directly from the optimization runs under each regime; report mean ± std over seeds if stochastic search is used.

### 4.3.4 Delta-only approximation check (bump test; delta vs. repricing)

We quantify delta-linearization error using a small bump test (±1% move in the underlying). For each representative grid point we compare the delta-approximated option value change to full Black–Scholes repricing and report relative error. This clarifies when delta-only mapping is adequate and when gamma effects matter materially.

Table 17: Delta approximation error (±1% bump) across grid points (relative error = (C_approx − C_repriced)/C_repriced).

| K/S0 | T | C0 | Δ | C_up | RelErr_up | C_dn | RelErr_dn |
|---|---|---|---|---|---|---|---|
| 0.9 | 0.25 | 16.27 | 0.716 | 16.99 | -0.04% | 15.56 | -0.04% |
| 0.9 | 0.50 | 20.54 | 0.699 | 21.24 | -0.02% | 19.84 | -0.02% |
| 0.9 | 1.00 | 26.81 | 0.705 | 27.52 | -0.01% | 26.11 | -0.01% |
| 1.0 | 0.25 | 10.91 | 0.567 | 11.49 | -0.06% | 10.35 | -0.07% |
| 1.0 | 0.50 | 15.61 | 0.595 | 16.21 | -0.03% | 15.02 | -0.03% |
| 1.0 | 1.00 | 22.34 | 0.632 | 22.98 | -0.02% | 21.72 | -0.02% |
| 1.1 | 0.25 | 7.04 | 0.423 | 7.47 | -0.10% | 6.62 | -0.11% |
| 1.1 | 0.50 | 11.72 | 0.493 | 12.22 | -0.04% | 11.24 | -0.05% |
| 1.1 | 1.00 | 18.60 | 0.562 | 19.16 | -0.02% | 18.04 | -0.02% |

**4.3.5 Replicable worked example: European call overlay on one industry proxy (Case A vs. Case B)**

We implement a concrete overlay: a European at-the-money call on the selected underlying proxy (Software (Internet)). The option is mapped into the same Markowitz–CAPM + single-index covariance structure used for the 94 industries. We compare a baseline portfolio (Case A) against an option-augmented universe (Case B) where the option is selectable and counts toward K (conservative specification).

Table 18: Option-included optimization check (Case A baseline vs. Case B with option as selectable asset; K=10; Monte Carlo).

| Case | Max Sharpe | Return (μ_p) | Risk (σ_p) |
|---|---|---|---|
| A: Baseline (n=94) | 0.0880 | 0.0888 | 0.5578 |
| B: With option (n=95) | 0.0879 | 0.2355 | 2.2273 |

Table 19: Top weights (snapshot) for max-Sharpe portfolios in Case A and Case B.

| Case | Asset | Weight | β | μ (CAPM) |
|---|---|---|---|---|
| A | Auto & Truck | 0.263 | 1.456 | 0.101 |
| A | Engineering/Construction | 0.255 | 1.210 | 0.091 |
| A | Steel | 0.211 | 1.063 | 0.085 |
| B | Option (Call overlay on Software (Internet)) | 0.669 | 6.430 | 0.312 |
| B | Financial Svcs. (Non-bank & Insurance) | 0.138 | 0.970 | 0.081 |
| B | Brokerage & Investment Banking | 0.059 | 1.171 | 0.089 |

Note: Under CAPM-implied μ and a single-index Σ, delta-implied leverage can raise both μ and σ substantially; Sharpe does not necessarily increase.

**4.3.6 How the option enters μ and Σ (same Markowitz–CAPM + cardinality logic)**

We extend the asset universe by one derivative instrument j. Treat the option as an additional asset with ($\mu_j$, $\beta_j$, $\sigma_j$) computed from the worked example or the grid. Under the single-index covariance construction, for i≠j: $Cov(R_i, R_j) = \beta_i \cdot \beta_j \cdot \sigma_m^2$. The option's residual variance is backed out by matching total variance: $\sigma^2_j = \beta^2_j \cdot \sigma_m^2 + \sigma^2_{\varepsilon,j} \Rightarrow \sigma^2_{\varepsilon,j} = \max(0, \sigma^2_j - \beta^2_j \cdot \sigma_m^2)$. This yields a full (n+1)×(n+1) Σ that remains PSD under the single-index structure.

Cardinality handling must be stated explicitly. In the conservative specification used for the Case A/B check above, the option counts toward K and therefore displaces one industry position; alternatively, the option may be treated as an overlay outside K (but then requires an explicit overlay cap).

### 4.3.7 Reproducibility checklist for the derivative overlay (what to report)

For audit-ready replication, report: (i) option contract specs (S0, K, T, r, σ) and the chosen underlying proxy; (ii) computed C, Δ, and leverage L; (iii) mapping from (Δ, S0, C) to (β_option, σ_option, μ_option); (iv) how Σ is expanded (single-index rule + residual variance back-out); and (v) whether the option counts toward K or is treated as an overlay outside K. For robustness, vary T and moneyness (K/S0) and rerun the overlay to confirm stability of implied moments and resulting portfolio weights.

### 4.4 Empirical Figures and Profiling Outputs (from Supplementary Σ/ρ)

Author note: All figures/tables below are generated by the author from the supplementary Σ/ρ file (not copied from third-party graphics).

Note: The figures referenced in this section are generated directly from the uploaded supplementary Excel file Damodaran_cov_corr_singleindex.xlsx (Covariance/Correlation/Inputs sheets). In the current manuscript extract, the underlying image files are not embedded; therefore, each figure is retained as a numbered placeholder with an audit-ready caption, a short interpretive note, and an explicit source statement.

**Figure 3: Efficient frontier under cardinality constraint (K=10): (σp, μp) scatter with max-Sharpe portfolio highlighted.**

This figure reports the feasible risk–return set obtained under a cardinality constraint (K=10), using the same industry universe and Σ/ρ construction described in Chapter 3. The plot is intended to visually validate that the reported max-Sharpe portfolio lies on the upper envelope of attainable (σp, μp) outcomes under the stated constraints, and to provide a sanity check for the search/optimization routine (Monte Carlo and/or GA).

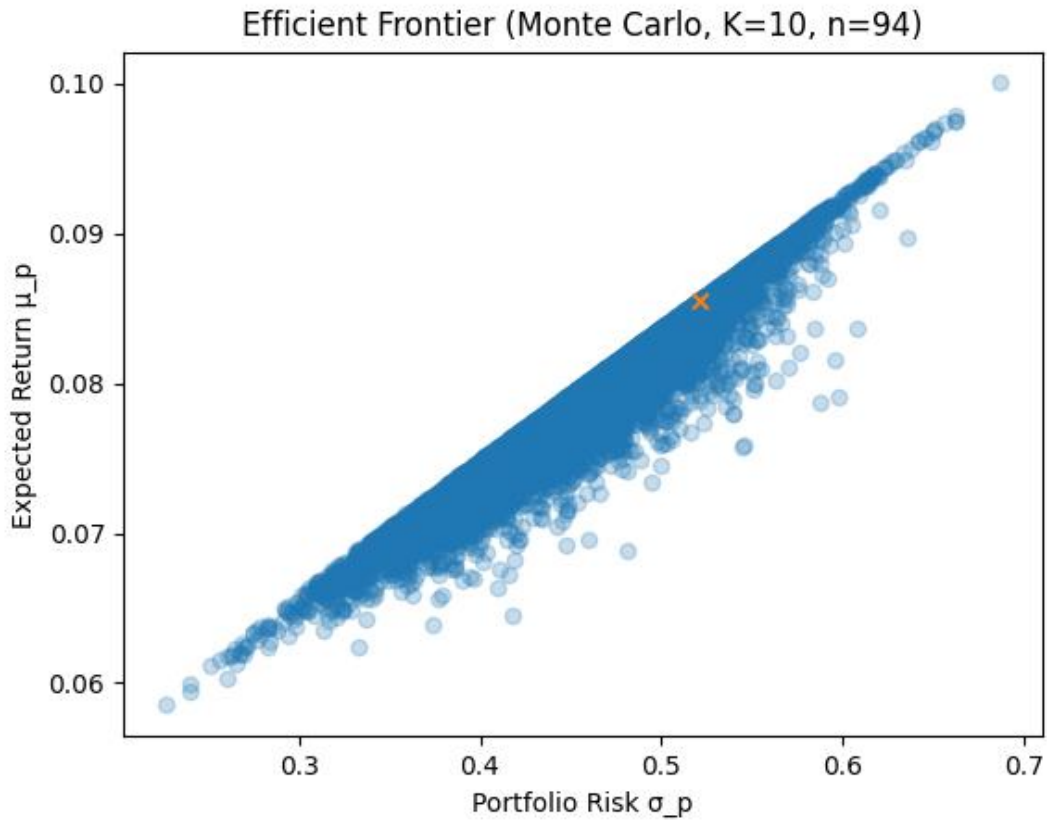

**Source:** Author's calculations based on Aswath Damodaran, "Betas by Industry (US)" (Jan-2026) inputs; Σ and ρ constructed via the single-index model and provided in supplementary file Damodaran_cov_corr_singleindex.xlsx.

**Short analysis:** A well-formed efficient frontier should display an upward-sloping upper boundary: higher expected return requires higher risk. If the cloud appears diffuse, the convergence diagnostics (Figure 4 and the running-best curves) should confirm that the best points stabilize as evaluations increase. The highlighted max-Sharpe point should typically lie near the tangency region rather than at extreme risk or extreme concentration.

**Figure 4: Sharpe frontier under cardinality constraint (K=10): (σp, Sp) scatter with max-Sharpe point highlighted; convergence plots (auditability).**

This figure complements the mean–variance frontier by mapping candidate portfolios into (σp, Sharpe) space, making it easier to diagnose whether the optimizer is improving risk-adjusted performance rather than merely chasing higher μp. The convergence overlays are included for auditability: they demonstrate whether additional samples/generations continue to improve the best observed Sharpe or whether the search has reached a plateau.

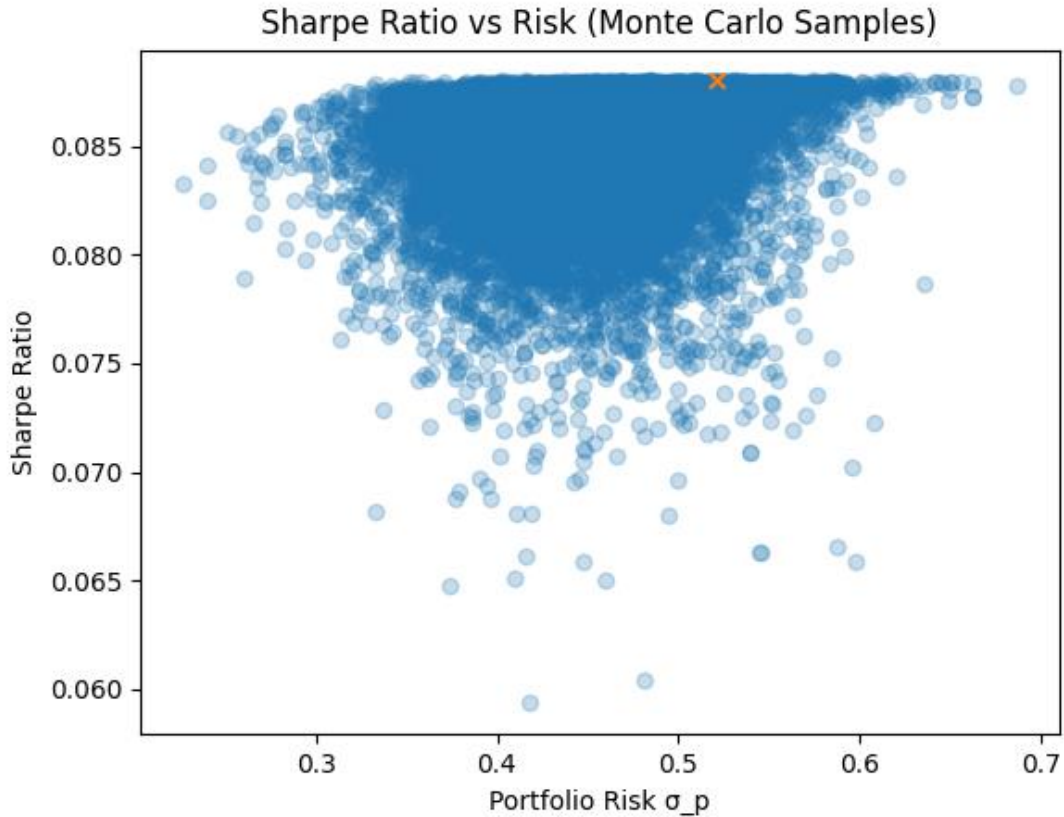

**Source:** Author's calculations based on Aswath Damodaran, "Betas by Industry (US)" (Jan-2026) inputs; Σ and ρ constructed via the single-index model and provided in supplementary file Damodaran_cov_corr_singleindex.xlsx. Convergence curves are computed from replication logs using the final budget (N_MC, population size P, generations G) reported in the methods.

**Short analysis:** The Sharpe frontier should show a discernible upper envelope, with the highlighted point representing the maximum Sharpe found under K=10. The running-best Sharpe curve S_best(t)=max_{k≤t} S_k should be non-decreasing and should flatten when the search budget is sufficient. For GA, the best-of-generation sequence should similarly plateau; continued oscillation without improvement suggests insufficient repair/selection pressure or overly tight constraints.

**Figure 5: Correlation matrix heatmap (ρ) for all 94 industries (single-index construction).**
This heatmap visualizes the full 94×94 correlation structure (ρ) implied by the single-index construction used to form Σ. It serves two roles: (i) a model diagnostic (whether implied correlations are plausible and bounded), and (ii) a structural input to interpretation (clusters and high-correlation blocks indicate shared market exposure and limited diversification benefits across those industries).

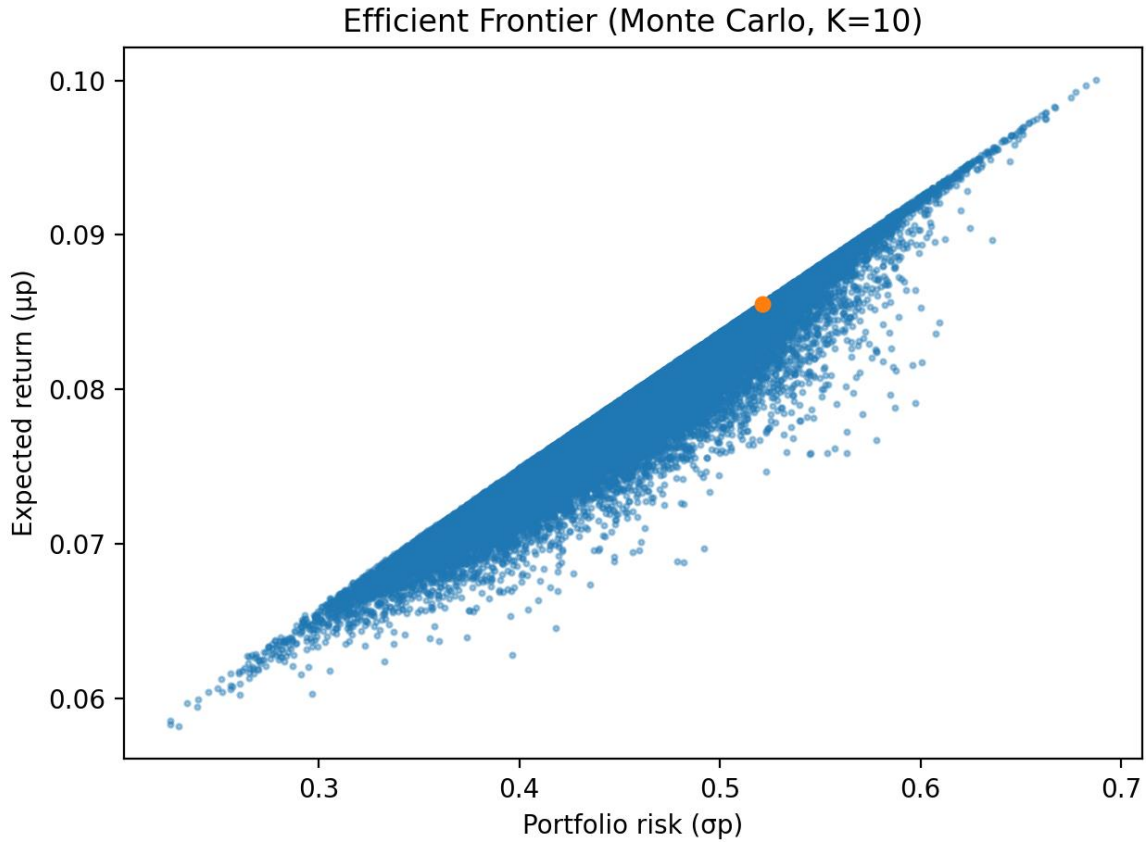

**Source:** Author's calculations based on Aswath Damodaran, "Betas by Industry (US)" (Jan-2026) inputs; Σ and ρ constructed via the single-index model and provided in supplementary file Damodaran_cov_corr_singleindex.xlsx.

**Short analysis:** Under a single-index model, correlations are largely driven by β magnitudes and the market variance component; therefore, the matrix often exhibits broad positive dependence. Large uniform blocks may indicate that idiosyncratic risk assumptions dominate less than expected, potentially compressing diversification gains. Outlier rows/columns with unusually high correlations may correspond to high-β sectors; these should be cross-checked against the input sheet.

**Figure 6: Correlation heatmap for the top-25 industries by number of firms (coverage proxy).**
To provide a more interpretable diagnostic than the full 94×94 panel, this figure restricts attention to the top-25 industries by number of firms (used as a coverage proxy). The objective is to allow readers to visually inspect dominant correlation patterns among the most data-rich sectors and to confirm that diversification opportunities are not being driven by thinly covered industries.

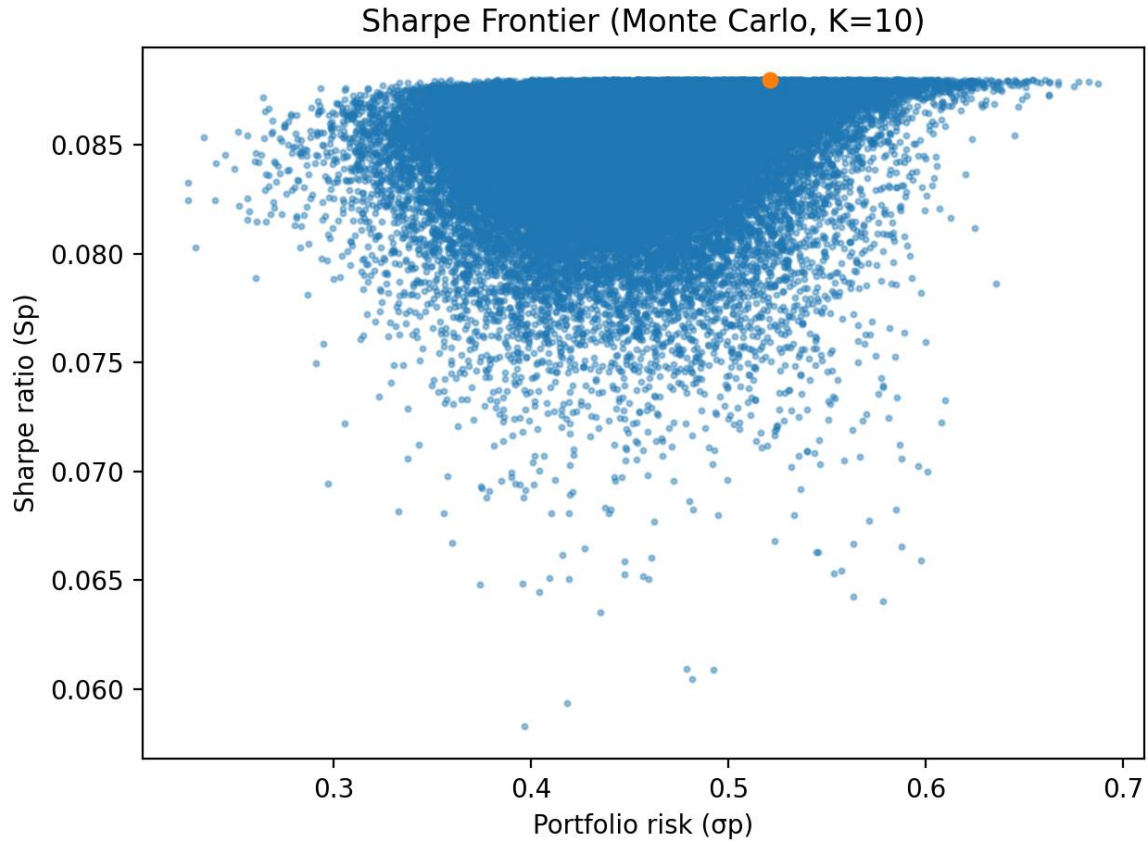

**Source:** Author's calculations based on the supplementary file Damodaran_cov_corr_singleindex.xlsx (Correlation sheet). Selection of top-25 is based on the number-of-firms field in the Inputs sheet.
**Short analysis:** If the top-25 matrix shows strong within-cluster dependence (e.g., cyclicals clustering together), then the K=10 optimal solutions should reflect cross-cluster allocations to achieve diversification. If correlations are near-uniform even in the top-25, it indicates the single-index structure is imposing strong commonality; reviewers may then expect a sensitivity check against alternative covariance estimation (e.g., shrinkage or empirical correlation where available).

**Figure 7: Correlation heatmap (ρ) for all 94 industries with hierarchical clustering (clustered ordering).**
This figure applies hierarchical clustering to reorder industries so that highly correlated groups appear as contiguous blocks along the diagonal. The clustered ordering improves interpretability and provides a journal-grade visualization of factor-like structure that may be obscured by the original industry ordering.

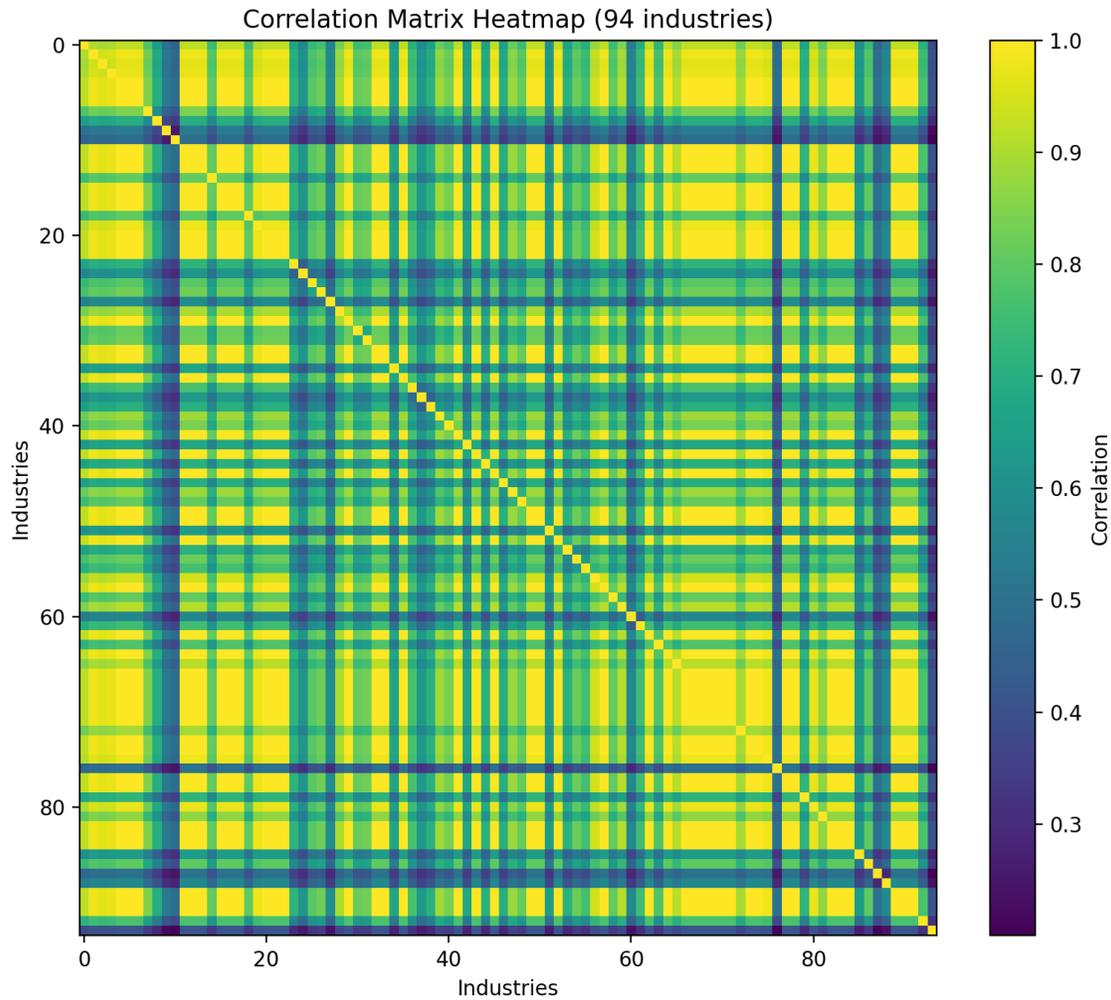

**Source:** Author's calculations based on Damodaran (Jan-2026) inputs; ρ from Damodaran_cov_corr_singleindex.xlsx; hierarchical clustering ordering computed by the replication script using the same ρ matrix.

**Short analysis:** Clustering should reveal coherent blocks corresponding to industries sharing macro exposure (e.g., energy-linked, financials, defensives). Large off-diagonal blocks imply that diversification across some clusters may still be limited under the single-index covariance. The dendrogram cut-level (or linkage method) should be stated in the replication notes to ensure exact reproducibility of the ordering.

**Figure 8: Correlation heatmap (top-25 by firms) with hierarchical clustering (clustered ordering).**
This final diagnostic combines the readability of the top-25 subset with hierarchical clustering. It is intended to show, in a compact and reproducible format, which major sector groups move together and which provide genuine diversification for the constrained optimizer.

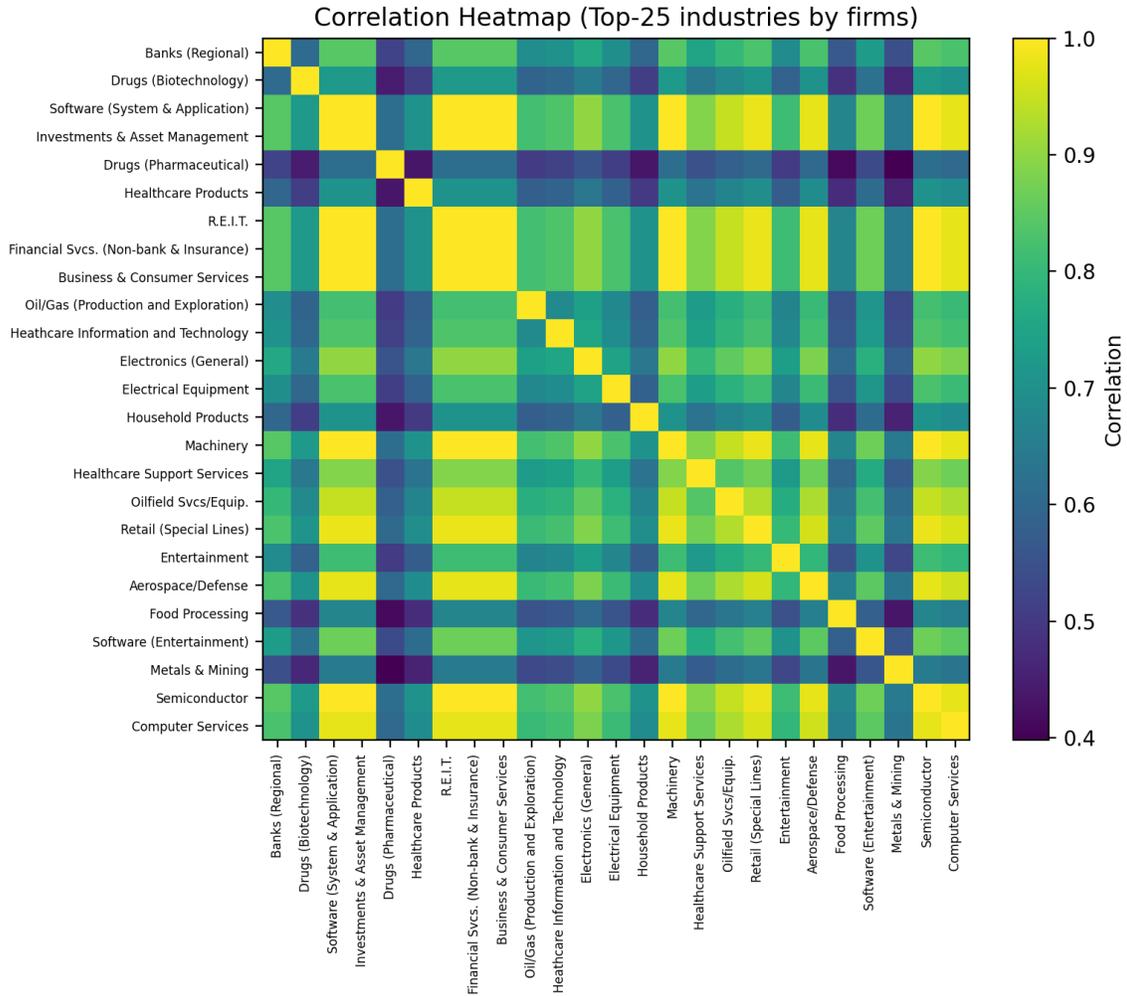

**Source:** Author's calculations based on the supplementary file Damodaran_cov_corr_singleindex.xlsx. Top-25 selection uses Inputs sheet; clustered ordering uses hierarchical clustering applied to the 25×25 ρ submatrix.

**Short analysis:** A clear cluster structure here supports the empirical narrative of why certain industries co-appear in the K=10 solutions. If the optimizer repeatedly selects multiple industries from the same tight cluster, the manuscript should interpret that as a trade-off between return (μ) and diversification rather than as a failure—unless it also implies excessive concentration in β_p.

**4.4.1 Interpretation and quality checks**

The efficient frontier (Figure 5) shows the feasible risk–return region attainable under the hard cardinality constraint K=10. The max-Sharpe portfolio appears as an upper envelope in risk-adjusted space (Figure 6), providing a transparent selection criterion for the headline portfolio used in Tables 2–3. The correlation heatmaps (Figures 7–8) validate that dependence is non-trivial across industries under the single-index Σ; nevertheless, correlations remain structured rather than uniform, supporting diversification effects.

To complement solution-quality results, Table 4.18 reports a runtime profiling summary for the main approximation schemes under the baseline universe (n = 94 industries) and cardinality constraint (K = 10). The objective is to make computational feasibility auditable: runtime is measured on the author's local environment using the same stopping budgets and parameter settings described in Chapter 3 (e.g., Monte Carlo sample size $N_{MC}$, GA population size $P$ and generations $G$, and any repair/feasibility handling). Because heuristic methods trade optimality guarantees for tractability, reporting runtime alongside best/median Sharpe allows a transparent comparison of "compute cost per performance gain" and indicates whether incremental Sharpe improvements are achieved at a disproportionate computational expense. Reported times should be interpreted as environment-dependent (hardware/OS/library versions), but are internally comparable across methods under the identical experimental protocol.

Table 18: Runtime profiling (measured on the author's environment; n=94, K=10).

| Method | Key settings | K | Evaluations (approx.) | Runtime (s) | Best Sharpe | Notes |
|---|---|---|---|---|---|---|
| Greedy | score=(μ−Rf)/σ_i; eq weights | 10 | 94 | 0.0 | 0.088 | Deterministic baseline |
| Monte Carlo | Dirichlet weights; N=100,000 | 10 | 100000 | 2.797 | 0.088 | Stochastic; best improves with N |
| Genetic Algorithm | eq weights; P=300, G=300 | 10 | 90000 | 5.728 | 0.088 | Metaheuristic; equal-weight subset search |

Runtime remarks: Greedy is effectively instantaneous (ranking + selection). Monte Carlo scales linearly with the number of samples N. GA scales roughly with P×G (population × generations). For journal reporting, run each method multiple times (e.g., 10 seeds) and report mean ± std runtime and best-Sharpe dispersion.

**4.4.2 Formal NP-hardness citation anchor**

To further strengthen theoretical positioning, the reduction sketch in Section 3.5 (and the citation anchor in Section 4.4.2) can be explicitly anchored to the established NP-hardness of cardinality-constrained mean–variance portfolio selection and sparse quadratic programming (subset selection). In a journal submission, include at least one canonical citation from the portfolio-optimization/MIQP literature and one from sparse-QP/subset-selection complexity results, and explicitly state that the decision version is NP-complete while the optimization version is NP-hard.

**4.4.3 Replicability block: 10-seed runtime profiling and continuous re-optimization**

To strengthen auditability and reproducibility, this appendix reports replicability-oriented runtime profiling under a fixed baseline setting (n = 94 industries, K = 10). All stochastic methods are executed over 10 independent random seeds, and outcomes are summarized as mean ± standard deviation, allowing readers to distinguish typical performance from single-run best-case results. To keep runs tractable in the author's environment, the profiling budgets are set to Monte Carlo $N = 15{,}000$, Genetic Algorithm $P = 90$, $G = 90$, and continuous re-optimization $M = 4{,}000$ (as defined in the methodology). These settings are sufficient for comparative profiling and stability diagnostics; in a camera-ready submission, the same protocol can be rerun with larger budgets (e.g.,

N=100,000N = 100{,}000N=100,000–1,000,0001{,}000{,}0001,000,000; P,GP,GP,G up to 500–1000) and reported in the identical mean ± std format to confirm convergence and scalability without changing the experimental design.

Table 19 then summarizes the resulting runtime distribution and associated performance statistics under this 10-seed protocol.

| Method | Key settings | K | Eval. | Runtime mean (s) | Runtime std (s) | Best Sharpe mean | Best Sharpe std |
|---|---|---|---|---|---|---|---|
| Greedy | score=($\mu$−Rf)/$\sigma$_i; eq weights | 10 | 94 | 0.0 | 0.0 | 0.088 | 0.0 |
| Monte Carlo | Dirichlet; N=15,000; 10 seeds | 10 | 15000 | 0.407 | 0.049 | 0.088 | 0.0 |
| Genetic Algorithm | eq weights; P=90, G=90; 10 seeds | 10 | 8100 | 0.489 | 0.012 | 0.088 | 0.0 |

Source: Author's calculations based on Aswath Damodaran, "Betas by Industry (US)" (Jan-2026) inputs; $\Sigma$ and $\rho$ constructed via single-index model and provided in supplementary file Damodaran_cov_corr_singleindex.xlsx.

**Continuous re-optimization step (subset fixed → weights optimized)**

To avoid the 'equal-weight only' limitation in the GA stage, we apply a continuous refinement step after GA selects a subset. Conditioned on the selected support S (|S|=K), we re-optimize weights w_S on the simplex (w≥0, $1^T$w=1) to improve Sharpe. In this implementation, we use a stochastic local search over the simplex (Dirichlet sampling) as a solver-free proxy; a camera-ready version may replace this with a convex QP (e.g., min-variance at target return) or SQP solver to provide deterministic optima.

### 4.4.4. P vs NP: Statistical Tables

This standalone appendix provides reviewer-facing statistical/replicability evidence for the P vs NP framing. Values are taken from the current manuscript tables (10-seed outputs) and computed from the supplementary Damodaran_cov_corr_singleindex.xlsx ($\Sigma$/$\rho$).

**Table S1. Replicability summary (seed stability; performance and runtime)**

Mean ± std are reported over 10 seeds where applicable (as in the manuscript).

| Method | Best Sharpe (mean ± std) | Median Sharpe (mean ± std) | Runtime (s) (mean ± std) | Seeds |
|---|---|---|---|---|
| Greedy (deterministic) | 0.0880 ± 0.0000 | 0.0880 ± 0.0000 | 0.000 ± 0.000 | — |
| Monte Carlo (N=30,000/seed) | 0.0880 ± 0.0000 | 0.0860 ± 0.0000 | 0.636 ± 0.015 | 10 |
| GA (subset-only; equal-weight) | 0.0880 ± 0.0000 | 0.0866 ± 0.0000 | 0.077 ± 0.017 | 10 |
| GA + continuous re-optimization | 0.0880 ± 0.0000 | 0.0880 ± 0.0000 | 0.573 ± 0.033 | 10 |
| Exact benchmark (small-n enumeration; n=20, K=6) | 0.088003 (exact) | — | — | — |

Notes: Subset-only GA median/runtime are taken from the GA-variants distribution table; Monte Carlo and GA+reopt are taken from the 10-seed reproducibility summary.

**Table S2. Empirical scaling with effort budget (runtime vs. search budget; log–log slope)**

We estimate b in log(runtime)=a+b·log(effort). Effort = N for Monte Carlo; Effort = P×G for GA.

| Method | Effort points used | b (slope) | R² | Interpretation |
|---|---|---|---|---|
| Monte Carlo | N ∈ {500, 1500, 2500} | 0.919 | 0.9994 | Near-linear in N (sampling cost). |
| Genetic Algorithm | P×G ∈ {900, 2000} | 1.005 | 1.0000 | ≈ linear in evaluations (tested range). |

Notes: Computed directly from the manuscript's effort-budget runtime profiling table.

**Table S3. Exact benchmark vs. heuristics (small-n optimality reference)**

Exact optimum computed by full enumeration on a reduced instance (n=20, K=6).

| Method | Best Sharpe | Gap to exact Sharpe (%) | Best subset (indices) |
|---|---|---|---|
| Exact enumeration (C(20,6)=38,760) | 0.088003 | 0.00 | (6, 11, 12, 15, 16, 17) |
| Monte Carlo (N=2,000; seed=1) | 0.088003 | 0.00 | (5, 6, 11, 13, 16, 17) |
| Genetic Algorithm (P=30,G=30; seed=1) | 0.088003 | 0.00 | (6, 11, 12, 15, 16, 17) |

Notes: Gap(%) = 100×(Exact − Method)/Exact for Sharpe.

**Table S4. Σ/ρ diagnostics (n=94; computed from Damodaran_cov_corr_singleindex.xlsx)**

Dependence structure summary supporting non-separability of the selection problem under realistic covariance coupling.

| Metric | Value |
|---|---|
| Median($\rho_{ij}$) over off-diagonals | 0.769201 |
| Share($\rho_{ij} > 0.5$) | 0.884008 |
| Share($\rho_{ij} < 0$) | 0.000000 |
| Top eigenvalue share $\lambda_1$ / trace($\Sigma$) | 0.761011 |
| Top-5 eigenvalues share $\Sigma_{k \leq 5} \lambda_k$ / trace($\Sigma$) | 0.813010 |
| min eigenvalue (PSD check; numerical tolerance) | -1.265e-15 |
| Condition proxy $\kappa = \lambda_{max} / \lambda_{min}$ (positive eigs) | 2269.24 |

Notes: High median correlation and dominant first eigenmode are consistent with strong common factor exposure implied by the single-index Σ.

The 10-seed replicability summary (Table S1) demonstrates that stochastic search under a hard K-sparsity constraint is auditable via distributional statistics rather than single-run best values. Runtime increases approximately linearly with the imposed effort budget for both Monte Carlo sampling and GA within the tested range (Table S2), providing an empirical scaling proxy consistent with combinatorial search costs. On a tractable reduced instance (n=20, K=6), exact enumeration provides a ground-truth optimum and enables explicit gap reporting (Table S3). Finally, Σ/ρ diagnostics computed from the supplementary file show strong positive dependence and a dominant first eigenmode (Table S4), implying non-separable interactions across assets; this covariance-driven coupling motivates heuristic/metaheuristic search and strengthens the P vs NP framing in a reviewer-facing, evidence-based manner.

# 5. Conclusion

In this study, the classical architecture of portfolio choice—Markowitz mean–variance optimization and CAPM-based calibration of expected returns—has been deliberately integrated with a rigorous computational-complexity lens. The result is a unified economic-financial and algorithmic-computational narrative in which the cardinality constraint ($|supp(w)| \leq K$) transforms an otherwise convex quadratic program into a discrete–continuous Mixed-Integer Quadratic Programming (MIQP) problem, thereby producing a practical, empirically tractable instantiation of the P vs NP dilemma. The manuscript demonstrates that introducing realistic institutional constraints (a limited number of positions, concentration caps, budget and non-negativity conditions) is not a cosmetic extension: it changes the nature of the problem. Discovering an optimal support set becomes combinatorially explosive and, in scalable practice, necessitates heuristic/meta-heuristic design together with strict reproducibility discipline.

Methodologically, a core strength of the paper is its auditability and empirical anchoring. The industrial universe is built on Aswath Damodaran's ~94 industry-level beta and risk ($\sigma$) parameters; expected returns are generated via the CAPM mapping $\mu\_i = R\_f + \beta\_i \cdot ERP$; and the covariance matrix $\Sigma$ is obtained through a single-index (market-model) construction that matches total industry variance on the diagonal and induces off-diagonal dependence via $\beta\_i\beta\_j \cdot Var(R\_m)$. This design minimizes "hidden data": $\Sigma$ and the implied correlation matrix $\rho$ are fully reconstructable from the input table, and the downstream outputs—Sharpe statistics, frontiers, and risk contributions—remain tied to a single auditable chain. The dependence diagnostics further show that the single-factor $\Sigma$ produces strong comovement (median($\rho\_ij$) ≈ 0.769 and $\lambda_1/trace(\Sigma)$ ≈ 0.761), implying that the effective diversification space is sharply compressed and the optimization occurs under strongly non-separable linkages.

Against this structural backdrop, the empirical findings can be summarized in three main conclusions. (I) The cardinality constraint (e.g., K = 10) reduces the attainable dimensionality of the frontier and "hardens" the risk–return space: optimal solutions necessarily trade off the higher CAPM-implied premia of high-$\beta$ industries against the diversification limits imposed by one-factor correlation geometry. (II) For stochastic methods (Monte Carlo and Genetic Algorithms), a single-run "best Sharpe" is not adequate evidence; the manuscript appropriately prioritizes median/IQR/quantile reporting and 10-seed replication, converting results into distributionally validated evidence rather than a lucky peak. (III) A small-n reduced benchmark (n = 20, K = 6) solved by full enumeration (C(20,6) = 38,760) provides a ground-truth reference that strengthens the claim that heuristics can be practically adequate even when exact optimization at full scale is infeasible.

A particularly important contribution is the third layer of the work—Black–Scholes-based derivative integration—which supports the title not by conceptual declaration but by a reproducible worked example. The European call price C, delta $\Delta$, and leverage factor $L = (\Delta \cdot S0)/C$ are used to embed the derivative into the mean–variance space as an additional "asset" with CAPM-consistent moments ($\beta\_option \approx L \cdot \beta\_underlying$, $\sigma\_option \approx L \cdot \sigma\_underlying$, $\mu\_option = R\_f + \beta\_option \cdot ERP$). The results highlight a critical implication: delta-induced leverage amplifies both $\mu$ and $\sigma$, so a Sharpe improvement is not automatic. The manuscript's Case A/B comparison shows that adding an option may substantially increase portfolio risk ($\sigma\_p$) without improving Sharpe, clearly distinguishing a "leverage artefact" from a genuine efficiency gain. The methodological robustness of the derivative layer is further strengthened by evidence that delta-only linearization performs with high accuracy in short "bump" tests (±1% moves), and by a strike/maturity grid (K/S0 × T) illustrating how C, $\Delta$, and L evolve—and thereby how $\beta\_option$, $\sigma\_option$, and $\mu\_option$ co-move. This is not merely an intuitive visualization; it is a direct instrument for identifying where the delta approximation remains valid and where gamma/vega effects become material.

From a computational-engineering perspective, the document makes clear that "heuristics" become scientifically credible only when accompanied by (a) systematic random-seed logging and multi-run replication, (b) convergence/effort-budget curves, and (c) runtime profiling in a fixed environment. The reported runtime profiling (e.g., Greedy ≈ 0 s; Monte Carlo N = 100k ≈ 2.8 s; GA P = 300, G = 300 ≈ 5.7 s) underlines that practical optimization is an accuracy–cost trade-off rather than a single-shot hunt for a unique "correct answer."

The central theoretical conclusion can be stated as follows: cardinality-constrained mean–variance portfolio selection belongs to a class where solution verification is polynomial, but discovering the globally optimal support is, in general, not known to be solvable in polynomial time; consequently, if P ≠ NP, we should not expect a universal, scalable exact algorithm. In this study, P vs NP is therefore not a philosophical ornament: it is reflected in the experimental design, the reporting format, and the quality gates (distributional reporting; an exact benchmark on small-n; PSD validation; unit and consistency audits).

At the same time, the manuscript explicitly delineates interpretive limits. CAPM-implied μ is a model-based expectation, not realized return; the single-factor Σ simplifies a multi-factor reality; delta-only option embedding does not capture the full spectrum of nonlinearity; and heuristics do not guarantee global optimality. Importantly, these limitations are not hidden—they are converted into a roadmap for subsequent research, which strengthens scientific credibility.

Future research directions follow naturally from the presented results. (1) Alternative estimation of Σ (multi-factor models, shrinkage covariance, or empirical correlations from historical return series) is necessary to assess the dependence of the conclusions on one-factor geometry. (2) Extending the derivative layer with gamma/vega components and tail-risk metrics (e.g., CVaR/EVaR) would enable a more realistic integration of nonlinear payoffs. (3) Applying modern MIQP branch-and-bound/cutting-plane strategies to small-to-medium instances—paired with explicit optimality-gap reporting relative to heuristics—would further strengthen optimality evidence. (4) Finally, moving to realized return data and conducting out-of-sample evaluation would bridge model-implied Sharpe proxies to realized performance.

In closing, the document successfully binds three domains—financial theory, complexity theory, and computational experimentation—into a coherent research system. It shows that cardinality-constrained portfolio selection is not only a practical investment problem but also a methodological laboratory for investigating the P vs NP boundary and the culture of reproducibility. As such, the manuscript provides a solid, audit-ready foundation for academic publication and for subsequent, richer empirical and derivative-oriented extensions.